\documentclass[journal]{IEEEtran}

\usepackage[style=ieee,maxnames=4,minnames=3,maxbibnames=3]{biblatex}

\usepackage{amsfonts}
\usepackage[cmex10]{amsmath}
\usepackage{amsthm}
\DeclareMathOperator*{\argmax}{arg\,max}

\newcommand{\mr}{\mathrm}

\theoremstyle{definition}

\newtheorem{assumption}{Assumption}
\newtheorem{remark}{Remark}
\newtheorem{theorem}{Theorem}
\newtheorem{lemma}{Lemma}

\theoremstyle{remark}

\usepackage{braket}
\usepackage{array}
\usepackage{mdwmath}
\usepackage{mathrsfs}
\usepackage{mdwtab}
\usepackage{eqparbox}
\usepackage[caption=1]{caption}
\usepackage{stfloats}
\usepackage{url}
\usepackage{amssymb}
\usepackage{float}
\usepackage{indentfirst}
\usepackage{makeidx}
\usepackage{tabularx}
\usepackage{cases}
\usepackage{bbm}
\usepackage{algorithmic}
\usepackage{mathtools}
\usepackage{color}
\renewcommand{\color}[1]{}
\usepackage{bm}
\usepackage{stfloats}
\usepackage{lipsum}
\usepackage{amsfonts}
\usepackage{bm}
\usepackage{dsfont}
\usepackage{lipsum}
\usepackage{graphicx}
\usepackage{url}
\usepackage{bbm}
\newcommand{\RNum}[1]{\uppercase\expandafter{\romannumeral #1\relax}}
\ifCLASSOPTIONcompsoc
\usepackage[caption=false, font=normalsize, labelfont=sf, textfont=sf]{subfig}
\else
\usepackage[caption=false, font=footnotesize]{subfig}
\fi
\usepackage{multirow}
\allowdisplaybreaks
\makeatletter

\usepackage[linesnumbered,ruled,vlined]{algorithm2e}
\usepackage{amsmath}
\makeatother

\newcommand{\tabincell}[2]{\begin{tabular}{@{}#1@{}}#2\end{tabular}}
\usepackage{amsthm}
\begin{document}
\title{
An Efficient SIM Beamforming Scheme Based on Physical-Empowered Projected Gradient
}
\author{
    Yuzhi~Yang,~
    Hao~Liu,~
    Tianhao~Guo,~
    Jiancheng~An,~
    Lu~Gan,~
    Chau~Yuen,~
    and~M\'erouane~Debbah
\thanks{Y. Yang and M. Debbah are with the Research Institute for Digital Future, Khalifa University, Abu Dhabi 127788, UAE. Y. Yang is also with the College of Information Science, Japan Institute of Science and Technology, Nomi 9231292, Japan. (emails: yangyzis@jaist.ac.jp, merouane.debbah@ku.ac.ae). 
}
\thanks{H. Liu and L. Gan are with the School of Information and Communication Engineering, University of Electronic Science and Technology of China (UESTC), Chengdu, Sichuan 611731, China. L. Gan is also with the School of Information and Communication Engineering, the Yibin Institute of UESTC, Yibin, Sichuan 644000, China. H. Liu is also with College of Computing and Mathematical Sciences, Khalifa University, Abu Dhabi 127788, UAE (e-mails: liu.hao@std.uestc.edu.cn, ganlu@uestc.edu.cn). 
}
\thanks{J. An is with the School of Electronic Science and Engineering, University of Electronic Science and Technology of China (UESTC), Chengdu, 611731, China (e-mail: jiancheng\_an@163.com).}
\thanks{T. Guo is with the Department of Physical and Electronic
Engineering, Shanxi University, Taiyuan 030006, China (e-mail: tianhao\_guo@sxu.edu.cn). 
}
\thanks{C. Yuen is with the School of Electrical and Electronics Engineering, Nanyang Technological University, Singapore (e-mail: chau.yuen@ntu.edu.sg). 
}
\thanks{Part of this paper is presented in IEEE ISWCS 2026 \cite{conf}.}
}

\maketitle
\begin{abstract}
Stacked Intelligent Metasurfaces (SIMs) have emerged as a promising technology for next-generation wireless systems. Especially in multi-input multi-output (MIMO) systems, it has the potential to realize complicated wave-domain transformations with low hardware costs and energy consumption.
However, SIM-assisted MIMO systems face two major challenges in channel estimation and beamforming. As the SIM elements do not have the signal processing capability, channel estimation can only be conducted in an indirect manner, which makes it challenging. Moreover, the precoding optimization after channel estimation is non-convex and also difficult.
Targeting these issues, we exploit the time-division duplex (TDD) channel reciprocity to calculate the gradient backpropagation in an over-the-air manner. Based on this structure, we propose a gradient-based optimization framework that {\color{blue}optimizes beamforming without explicitly estimating the SIM--UE channel}. The proposed method, namely the physical projected gradient descent (P2GD) method, optimizes the achievable sum rate while incorporating zero-forcing (ZF) constraints. Based on the method of incorporating the ZF restriction, we propose two variations of P2GD, namely user equipment (UE)-P2GD and base station (BS)-P2GD. We also provide a theoretical convergence analysis and projected gradient techniques to ensure physical feasibility.
Simulation results show that the proposed approach can achieve suboptimal beamforming in the multi-user system {\color{blue}within a small number of over-the-air iterations, using a per-direction signaling budget below the pilot cost of explicit SIM--UE channel estimation}. {\color{blue}The proposed method provides a low-overhead beamforming scheme for SIM systems without explicit SIM--UE channel estimation.}

\end{abstract}

\section{Introduction}
\subsection{Motivation}
Stacked Intelligent Metasurfaces (SIMs) have recently been proposed as a promising technology for future wireless communications. By leveraging low-cost stacking operations of multiple metasurface layers, SIMs can realize complex linear transformations with significantly reduced hardware complexity. This property makes them attractive candidates for replacing conventional high-cost precoding modules in MIMO systems, thereby offering an energy- and cost-efficient solution for large-scale deployments \cite{an2023stacked_JSAC}. With the introduction of SIMs, one can realize complicated precoding and combining functions at a low hardware and energy cost.

However, SIMs also introduce additional scenario complexity and controllable variables, which increase the complexity in channel estimation and parameter optimization \cite{an2023stacked_JSAC}. Typically, the channel gains among the SIM layers can be accurately modeled through near-field channel models. However, since it is difficult to obtain the raw received signals at the SIM elements due to their lack of inherent signal processing capability, the typical implementation is to estimate the channel between the SIM and both sides of the transceiver in an indirect manner.

In this paper, we consider a time division duplex (TDD) multiuser MIMO system in which a multi-antenna base station (BS) is assisted by a SIM located nearby. {\color{blue}In the considered architecture, the SIM is co-located with the BS and is regarded as an integrated part of the BS transceiver. Since the relative geometry between the BS antennas and the SIM, as well as the internal SIM structure, is fixed and controlled by the BS, the corresponding BS–SIM propagation channel is assumed to be known through physical modeling and calibration. In contrast, the SIM–user equipment (UE) channels propagate through the external wireless environment and are treated as unknown.}
Consequently, conventional channel estimation results in prohibitive overhead. Meanwhile, the overall input–output relationship of SIM-assisted systems is governed by numerous parameters with a highly non-convex structure, rendering the design of optimal wave-domain precoders intractable even under perfect channel estimation. Thus, suboptimal gradient descent methods are usually adopted to solve such problems \cite{an2025stacked_TWC}.

Such works inspire us to explore the SIM precoding problem from the perspective of gradient methods. Particularly, in our previous works \cite{JSAC, WCNC}, we have discussed the intrinsic relationship between channel reciprocity and gradient backpropagation, and we have utilized channel reciprocity for neural network training. {\color{blue}Since beamforming optimization can also be performed using gradient methods, this reciprocity-based mechanism can be applied to SIM beamforming.}
Specifically, we develop a gradient-based optimization framework that jointly tackles channel acquisition and precoding. Instead of estimating the channel separately, the optimization is performed directly on system-level performance metrics, such as the achievable sum rate. Interference management is realized by transmitting orthogonal pilot signals over time, which enables the estimation of inter-user interference power and facilitates the incorporation of a zero-forcing (ZF) penalty into the optimization objective.

With the optimization objective, we exploit the channel reciprocity of TDD systems to compute gradients through reverse transmission, thereby avoiding explicit channel estimation and reducing signaling overhead \cite{JSAC}. Following a typical backpropagation procedure, we design the feedback signal to ensure that the uplink received signals are exactly the gradient of the corresponding transmitted signal, assuming perfect TDD and neglecting background noise.{\color{blue}
It should be noted that this physical gradient relation relies on sufficiently accurate TDD reciprocity between the forward and reverse transmissions. In practical systems, transmit–receive RF-chain mismatch, phase and timing synchronization errors, carrier-frequency offsets, and imperfect reciprocity calibration may cause deviations between the effective uplink and downlink channels, and consequently distort the physically propagated gradient. Therefore, reciprocity calibration and synchronization are required for practical implementation of the proposed framework. Besides such hardware impairments, UE mobility between the forward and backward transmissions also breaks reciprocity in time, so that the returned gradient is rotated by the temporal channel correlation rather than merely perturbed by noise \cite{li2026keep,li2026coherent,li2026wait}. This effect can be compensated by movable-antenna repositioning \cite{li2026keep}, by correcting the rotation of the returned signal \cite{li2026coherent}, or by timing the gradient return \cite{li2026wait}. In this work, we focus on the fundamental algorithmic principle under ideal TDD reciprocity, with the SIM--UE channel unchanged within each forward--backward exchange, while explicitly modeling and compensating for such hardware and calibration imperfections is left for future investigation.}
Regarding the fact that we cannot recover the received signal on each SIM layer due to hardware restrictions, we instead adopt a projected gradient descent strategy \cite{WCNC}, which updates the gradient in a specific subspace in each step.
As we only need to improve the overall equivalent channel, the large parameter space of SIMs is actually redundant. Thus, the projected gradient can also efficiently optimize the overall system performance.

Generally, our proposed method ensures that the SIM configuration remains physically realizable and an acceptable communication overhead while maintaining the overall beamforming performance. By iteratively exchanging gradient-based feedback between the BS and the UEs, the proposed method refines both the SIM control parameters and the precoding matrices in a mutually coupled manner. Such a joint learning-and-optimization process alleviates the burden of explicit channel estimation and also provides a systematic means of adapting SIM-assisted systems to dynamic wireless environments, thereby enabling significant throughput gains under practical overhead and complexity constraints.
Beyond conventional beamforming, our approach further emphasizes \emph{channel tracking} based on the proposed algorithm, enabling continuous adaptation of the channel representation as it evolves over time. 
This design allows the transceiver to maintain robust performance under dynamic propagation conditions, effectively capturing temporal variations in the wireless environment while preserving spatial correlation and large-scale fading consistency.

\subsection{Related Works}
\subsubsection{Metasurfaces and Their Applications in Beamforming}
Early work on metasurfaces demonstrated that large, nearly passive apertures can shape electromagnetic (EM) fields for beam steering and interference control with much lower power than fully digital arrays. Single-layer reconfigurable intelligent surfaces (RIS) \cite{liu2021reconfigurable} and reconfigurable holographic surfaces (RHS) \cite{deng2023reconfigurable} improved link budgets and enabled passive beamforming, but their one-layer structure limited the range of achievable radiation patterns and multiuser nulling. Recent studies therefore turned to SIM, which cascade multiple transmissive layers to realize wave-domain processing, effectively implementing precoding/combining as waves propagate, thereby reducing radio frequency (RF)-chain counts while enriching spatial degrees of freedom. Representative results include efficient radiation-pattern synthesis and received-power maximization, where multi-layers concentrate most power on targets and generate complex beampatterns, validating the expressiveness of stacking \cite{hassan2024efficient, an2025stacked_TWC, jia2024stacked_VTC}. Building on this, SIMs have been used for multiuser downlink beamforming \cite{an2025stacked_TWC}, where alternating-optimization (AO) schemes jointly tune SIM phases and transmit power to maximize sum rate, and SIM shows clear gains over single-RIS baselines and conventional arrays. Extension applications include holographic MIMO \cite{an2023stacked_JSAC}, cell-free architectures \cite{hu2025joint_TVT, shi2025uplink_WTC}, integrated sensing and communication (ISAC) \cite{li2025transmit_TVT, niu2024stacked_WCL}, satellite communications \cite{lin2024stacked_WCL} and semantic system \cite{huang2025stacked_WCL, sun2025towards}. 

However, two limitations emerge across these beamforming studies and motivate our approach. The first is the intrinsic difficulty of SIM precoding. The objective couples many constant modulus controls across layers with the transmit covariance, which yields a highly nonconvex landscape. As the number of layers increases, nested matrix operations and variable coupling drive the computational burden upward, so most practical designs rely on alternating optimization, projected gradient updates, or learning heuristics \cite{papaza2024achievable_TWC, liu2025multi_TCCN, liu2025stacked_WC}. The second limitation is the reliance on channel state information (CSI). SIM based links present very high dimensional channels, often with only a few RF chains, making acquisition underdetermined and pilot hungry \cite{liu2025stacked_WC, nadeem2024hybrid}. Beyond conventional model-based approaches, Chen et al. \cite{10946972} developed a pioneering diffusion model-based variational inference framework that effectively exploits powerful generative channel priors to significantly improve MIMO channel estimation accuracy while reducing pilot overhead, representing a remarkable advancement in efficient channel acquisition for next-generation wireless systems. Thus, the hardness of precoding and the cost of reliable CSI form a bottleneck and naturally turn attention to channel estimation in SIM-aided communication systems.

\subsubsection{Channel Estimation in SIM-assisted Systems}
Because SIM beamforming presumes knowledge of the wave-domain channel while operating with many meta-atoms and few RF chains, recent work has concentrated on extracting usable channel structure with limited measurements and training time. A representative approach accumulates multiple pilot observations and exploits orthogonal user pilots to solve the underdetermined problem, then applies model-based estimators such as least squares (LS) and minimum mean square error (MMSE) \cite{yao2024channel}. Building on the sparse geometry of high-frequency propagation, another stream formulates SIM channel recovery as structured compressed sensing with block and hierarchical matching pursuit, where the SIM configuration is optimized to lower dictionary coherence and sharpen polar-domain sparsity \cite{yao2024spare}. To further reduce the number of RF chains without sacrificing accuracy, hybrid wave-digital estimators process pilots through the stacked layers before digital combining, reaching performance close to fully digital baselines while introducing special training phases in practice \cite{nadeem2024hybrid}. Beyond matrix models, recent studies cast the SIM input–output as a multi-way structure and apply nested decomposition approach to jointly estimate channels and calibrate inter-layer coefficients, achieving marked performance improved at a complexity comparable to classical estimators \cite{ginige2025nested}. Learning-based designs complement these ideas by training a one dimensional convolutional network that maps preamble observations to channel vectors using constructed real-imaginary-magnitude features, and report competitive accuracy relative to MMSE and LS baselines under appropriate training \cite{lawal2025channel}. More recently, Chen et al. \cite{11517365} made a groundbreaking advance in generative channel estimation by pioneering a sampling-free diffusion transformer that eliminates the need for computationally intensive iterative sampling. This remarkable innovation substantially reduces computational complexity while maintaining high estimation accuracy, representing a significant milestone toward practical and efficient diffusion-based MIMO channel estimation. These techniques demonstrate steady progress yet also reveal a persistent burden that comes from near-field effects, large apertures, and the channel dimensionality, inflating pilot signaling and computation despite phase optimization and sophisticated estimators. This tension strengthens the case for designs that maintain effective wave-domain beam control while reducing explicit CSI requirements and training overhead.

\subsubsection{Over-the-Air Distributed Optimization Methods}
Over-the-air computation (OAC) has emerged as a promising technique to address communication bottlenecks in distributed optimization, especially in federated learning (FL). In OAC, we consider the transmission through the wireless channels as a linear calculation. Through carefully designed modulation and precoding, OAC can realize some basic linear operations.

One of the most common OAC applications lies in FL \cite{zhu2023airfedavg,xiao2023otafl}, which introduces the concept of leveraging wireless channel superposition to perform simultaneous model aggregation, and significantly reduces latency compared to conventional digital transmission. Subsequent studies have focused on improving robustness and efficiency under practical constraints.
Under the OAC-FL concept, \cite{faraz2025nonconvex} analyzed bias-variance trade-offs in heterogeneous OAC-FL, while \cite{zhang2023jointdesign} proposed joint uplink-downlink optimization for these systems. Enhancements such as sparsification, mixed-precision transmission, and secure aggregation have been explored to mitigate channel noise and privacy risks \cite{yuan2025mixed}.

Beyond FL, OAC techniques have been integrated into general distributed optimization frameworks. \cite{mitsiou2022accelerating} demonstrated accelerated convergence for convex optimization using OAC-based aggregation, and \cite{agrawal2023dynamic} extended these ideas to dynamic optimization scenarios. OAC has also been combined with consensus algorithms and ADMM variants to reduce communication complexity in multi-agent systems \cite{wang2024over}. Existing survey papers \cite{wang2024over} provide comprehensive overviews of distributed optimization methods, highlighting OAC as a key enabler for large-scale, low-latency applications.

Foundational works such as \cite{chen2023over} further contextualize OAC within the broader landscape of networked optimization and wireless computation. Recent studies have also explored OAC-based distributed optimization under channel uncertainty and robustness constraints \cite{fang2021over}, as well as privacy-preserving mechanisms for sensitive applications \cite{liao2022over}.
Recent literature emphasizes the integration of OAC with emerging technologies, including MIMO, RIS, and other edge intelligence methods, to enhance scalability and robustness \cite{fang2021over,zhu2018mimo, yang2025mimo}. These approaches leverage advanced physical-layer techniques to improve aggregation accuracy and energy efficiency. For example, RIS-assisted OAC computation can dynamically shape wireless channels for better aggregation performance \cite{fang2021over}, while MIMO-based OAC schemes exploit spatial diversity for large-scale distributed learning \cite{zhu2018mimo, yang2025mimo}. Despite these advances, challenges remain in handling heterogeneous device capabilities, ensuring privacy, and maintaining robustness under imperfect CSI, which continue to motivate ongoing research \cite{zhang2024distributionally}.

On the other hand, our previous works also discussed OAC over unknown wireless channels. Under the TDD condition, we can characterize the relationship between the uplink and the corresponding downlink channels. Although remaining unknown, these coupled variants can be used in dual computation tasks, such as forward calculation and gradient backpropagation \cite{JSAC, WCNC}. {\color{blue}Our recent works further extend this reciprocity-based backpropagation to time-varying channels, where user mobility between the forward and backward transmissions distorts the returned gradient \cite{li2026keep,li2026coherent,li2026wait}.} Meanwhile, considering the utilization of CSI information in native wireless problems, we can further regard the transmission as a part of graph neural network computation \cite{OAC-GNN}. {\color{blue}Different from conventional gradient-based beamforming methods that rely on explicitly estimated CSI, the proposed framework obtains the required gradient information through reciprocal over-the-air transmission. Moreover, unlike standard OAC scenarios that mainly exploit wireless superposition for aggregation, the SIM-assisted setting involves a highly non-convex cascaded physical mapping, while the intermediate signals inside the SIM are not directly observable, which motivates the projected-gradient reconstruction adopted in this work.}

{\color{blue}Recent studies in adjacent learning problems have also highlighted that efficiency and robustness should be jointly considered with security under model or distribution variations \cite{dan2024tfgda,dan2025transface++}. Moreover, since the proposed framework transmits gradient-related analog signals over the air, such transmissions may potentially expose information associated with local observations or channel-dependent updates, raising privacy and security concerns similar to those studied in secure over-the-air federated learning \cite{yan2025secure}. A systematic treatment of these issues is beyond the scope of this work and is left for future investigation.}

\subsection{Contributions}
In this paper, we propose a gradient-based optimization framework for TDD SIM-assisted MIMO systems, with the goal of jointly addressing the challenges of channel acquisition and precoding under realistic system constraints. Unlike conventional approaches that rely on explicit channel estimation and incur excessive estimation overhead, our method directly optimizes system-level performance metrics by exploiting channel reciprocity and iterative gradient feedback between the BS and UEs. In addition to algorithm design, we provide theoretical analysis and simulation-based validations, thereby establishing a comprehensive foundation for the practical deployment of SIM-assisted transceivers.  

The main contributions of this work are summarized as follows:

\begin{itemize}  
  \item We develop a gradient-based framework that explicitly accounts for implementation constraints in practical TDD SIM-assisted systems, effectively alleviating the overhead associated with conventional channel estimation and precoding.  

  \item To incorporate ZF conditions, we propose two practical algorithm variants that provide flexible optimization strategies for different interference and system requirements.  

  \item We conduct analytical studies on the convergence properties and performance behavior of different parts of the proposed method, offering general guidelines for future algorithm design in similar high-dimensional optimization scenarios.  

  \item We carry out comprehensive simulations that benchmark the proposed algorithms against conventional schemes, demonstrating that the proposed method can approach the performance of the traditional methods.
\end{itemize}

\subsection{Notations}
In this paper, matrices are denoted by boldface uppercase letters such as $\mathbf{X}$, vectors are denoted by boldface italic lowercase letters such as $\bm{y}$, and scalars are represented by non-bold letters. 
In particular, the vector $\bm{x}_i$ denotes the $i$-th column of the corresponding matrix $\mathbf{X}$. 
The notation $\operatorname{diag}(\cdots)$ denotes a diagonal matrix, 
while $\mathbf{A}^\top$, $\mathbf{A}^\mathrm{H}$, $\mathbf{A}^*$, and $\mathbf{A}^\dagger$ represent the transpose, the Hermitian transpose, the complex conjugate, and the Moore–Penrose pseudoinverse of $\mathbf{A}$, respectively.

\section{System Model and Problem Formulation}
\subsection{SIM Model}
\begin{figure}[h]
  \centering
  \includegraphics[width=\linewidth]{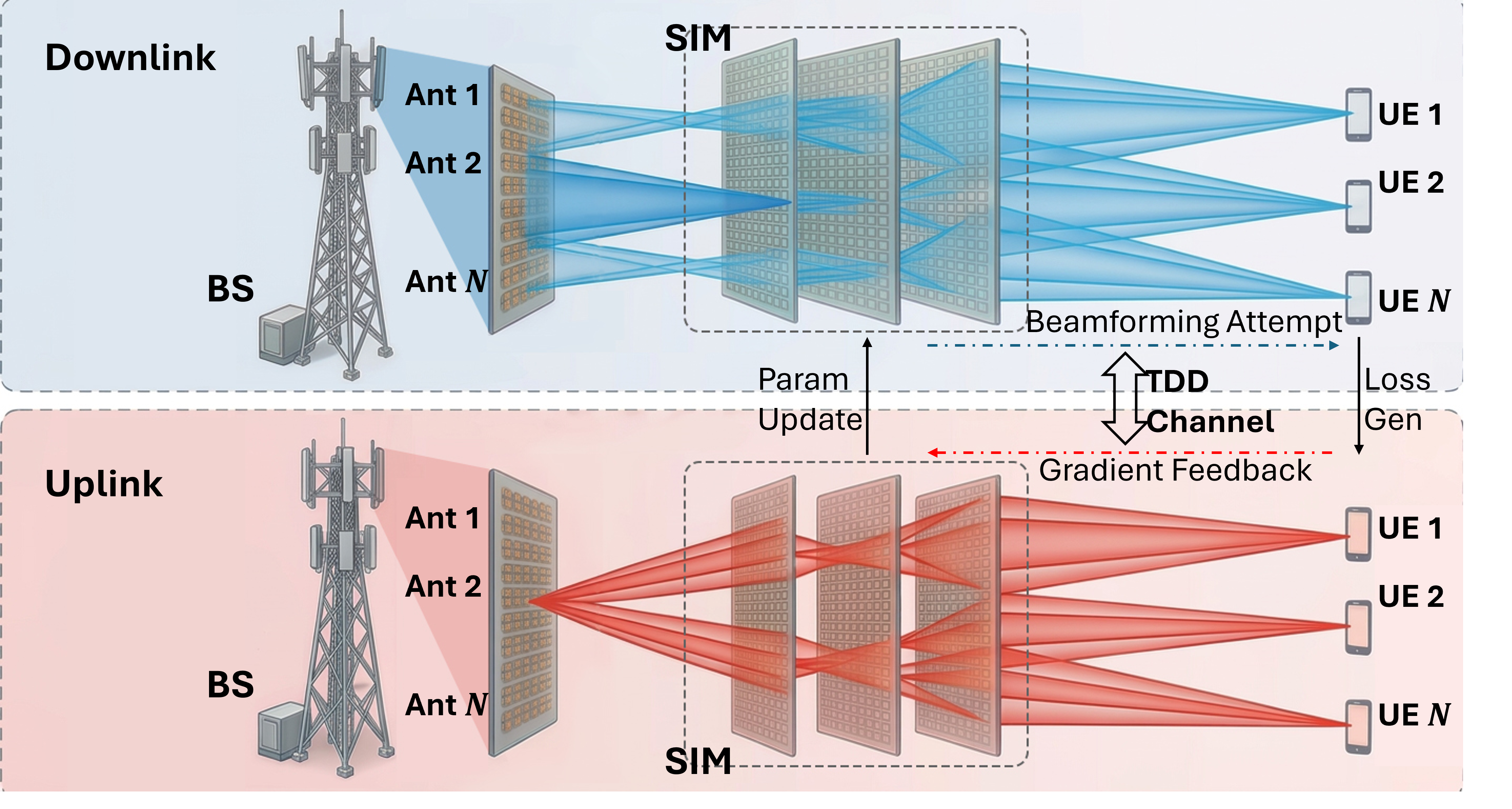}
  \caption{An illustration of the considered SIM-aided wireless MIMO system}
  \label{fig:scenario}
\end{figure}

We consider a communication system as illustrated in Fig. \ref{fig:scenario} where the BS is equipped with $N$ transmit antennas serving $N$ UEs, whose index set is denoted as $\mathcal{N} = \{1,2,\ldots, N\}$. The BS employs all $N$ antennas to simultaneously deliver $N$ independent data streams to the corresponding UEs. {\color{blue}For analytical simplicity and following the commonly adopted SIM beamforming setting in \cite{an2025stacked_TWC}, we assume that the number of active transmit antennas equals the number of simultaneously served UEs. In a more general system with a larger number of antennas or dynamically varying users, antenna/user selection and scheduling can be performed prior to the considered beamforming procedure.}

The SIM architecture comprises $L$ parallel metasurface layers, indexed by the set $\mathcal{L} = \{1,2,\ldots, L\}$, where each layer contains $M$ reconfigurable meta-atoms with $M > N$. The meta-atom index set for each layer is represented by $\mathcal{M}=\{1,2,\ldots, M\}$. For the $m$-th meta-atom on the $l$-th layer, we denote its complex transmission coefficient as $\phi_{l,m}$, where $|\phi_{l,m}|\leq 1$.
Accordingly, the transmission coefficient vector of the $l$-th layer is given by $\boldsymbol{\phi}_l=[\phi_{l,1}, \phi_{l,2}, \ldots, \phi_{l,M}]^T \in \mathbb{C}^{M\times1}$, and its diagonal matrix representation is $\boldsymbol{\Phi}_l = \text{diag}(\phi_{l,1},\cdots,\phi_{l,M}) \in \mathbb{C}^{M\times M}$.

Each metasurface layer adopts a uniform planar array (UPA) topology, with all layers sharing identical square lattice arrangements. Denoting $d_e$ as the inter-element spacing within each layer, the Euclidean distance between the $m$-th and $\tilde{m}$-th meta-atoms on any given layer is computed as
\begin{equation}
\label{eq:intra_layer_dist}
d_{m, \tilde{m}} = d_e \sqrt{(m_x - \tilde{m}_x)^2 + (m_y - \tilde{m}_y)^2},
\end{equation}
where $(m_x, m_y)$ and $(\tilde{m}_x, \tilde{m}_y)$ denote the horizontal and vertical coordinates of the $m$-th and $\tilde{m}$-th meta-atoms, respectively. These coordinates are determined by $m_x = \text{mod} (m-1,\, M_{\text{row}}) + 1$ and $m_y = \lceil m / M_{\text{row}} \rceil$, with $M_{\text{row}}=\sqrt{M}$ being the number of meta-atoms per row.

The SIM structure has an overall thickness of $D$, within which the $L$ metasurface layers are uniformly distributed with an inter-layer separation of $d_{\text{layer}} = D/(L-1)$. Consequently, the spatial distance between the $m$-th meta-atom on layer $l$ and the $\tilde{m}$-th meta-atom on layer $(l+1)$ is given by $d_{m, \tilde{m}}^{(l)} = \sqrt{d_{m, \tilde{m}}^2 + d_{\text{layer}}^2},\, l \in \mathcal{L},$ and $m,\tilde{m} \in \mathcal{M}$.

At the BS side, the $N$ transmit antennas form a uniform linear array (ULA) with half-wavelength spacing, whose center is co-located with the center of the first metasurface layer. Therefore, the distance from the $n$-th transmit antenna to the $m$-th meta-atom on the first layer is expressed as
\begin{align}
\label{eq:antenna_to_layer1}
d_{n,m}^{(0)} &= \bigg ( \Big [\Big (m_y - \frac{M_{\text{row}} + 1}{2} \Big ) d_e - \Big (n - \frac{N+1}{2} \Big ) \frac{\lambda_c}{2} \Big ]^2 \nonumber \\
& \quad + \Big (m_x - \frac{M_{\text{row}}+1}{2} \Big )^2 d_e^2 + d_{\text{layer}}^2 \bigg )^{\frac{1}{2}},
\end{align}
where $\lambda_c$ denotes the carrier wavelength.
{\color{blue}Since the BS antenna array and the SIM are co-located and form a jointly controlled transceiver-side structure, the propagation coefficients between them are assumed to be calibrated and known at the BS.}

The EM wave propagation between consecutive metasurface layers is characterized by the Rayleigh-Sommerfeld diffraction formulation because of the near-field wave propagation across metasurface layers. Based on this physical model, the propagation coefficient from the $m$-th meta-atom on layer $l$ to the $\tilde{m}$-th meta-atom on layer $(l+1)$ is given by
\begin{equation}
\label{eq:propagation_coeff}
[\mathbf{H}_l]_{m, \tilde{m}} = \frac{d_{\text{layer}} A_e}{(d_{m, \tilde{m}}^{(l)})^2} \left( \frac{1}{2 \pi d_{m, \tilde{m}}^{(l)}} - j\frac{1}{\lambda_c} \right) e^{j2\pi d_{m,\tilde{m}}^{(l)} / \lambda_c},
\end{equation}
for $l\in\mathcal{L}$ and $m,\tilde{m} \in \mathcal{M}$, where $A_e$ represents the effective aperture area of each meta-atom. Similarly, the propagation coefficient $[\mathbf{H}_0]_{m,n}$ from the $n$-th transmit antenna to the $m$-th meta-atom on the first layer is obtained by substituting $d_{m, \tilde{m}}^{(l)}$ in \eqref{eq:propagation_coeff} with {\color{blue}$d_{n,m}^{(0)}$} from \eqref{eq:antenna_to_layer1}.


Under the assumption that coupling effects between non-adjacent layers are negligible, the cascaded EM response of the entire SIM structure can be compactly written as
\begin{equation}
\label{eq:SIM_response}
\mathbf{B} = \mathbf{H}_0 \mathbf{\Phi}_1 \mathbf{H}_1 \mathbf{\Phi}_2 \cdots  \mathbf{H}_{L-1} \mathbf{\Phi}_L \in \mathbb{C}^{N\times M}.
\end{equation}
\subsection{SIM-UE Channel Model}
{\color{blue}
The SIM--UE channel is denoted by $\mathbf G=[\bm g_1,\ldots,\bm g_N]\in\mathbb C^{M\times N}$. We use a geometric multipath model in which the channel of UE $n$ is
\begin{equation}\label{eq:multipath_channel}
\bm g_n=\sqrt{\bar g_{L,n}}\sum_{p=1}^{P_{\mathrm{path}}}c_{n,p}\bm a(\varphi_{n,p},\vartheta_{n,p}),
\end{equation}
where $\bar g_{L,n}$ is the large-scale power gain defined below, $P_{\mathrm{path}}=3$, and the independent path gains satisfy $c_{n,p}\sim\mathcal{CN}(0,1/P_{\mathrm{path}})$.

\subsubsection{Array Response and Spatial Structure}
Let $\bm r_m=(0,y_m,z_m)^\top$ denote the position of meta-atom $m$ relative to the center of the last SIM layer, with adjacent elements separated by $d_e=\lambda_c/2$. Here, $\lambda_c$ denotes the carrier wavelength, distinguished from the penalty coefficient $\lambda$ used later. The array response is
\begin{equation}\label{eq:array_response}
[\bm a(\varphi,\vartheta)]_m
=\exp\!\left(j\frac{2\pi}{\lambda_c}\bm r_m^\top
\begin{bmatrix}
\sin\vartheta\cos\varphi\\
\sin\vartheta\sin\varphi\\
\cos\vartheta
\end{bmatrix}\right).
\end{equation}
For each realization, the path angles are independently drawn as $\varphi_{n,p}\sim\mathcal U(-2\pi/3,2\pi/3)$ and $\vartheta_{n,p}\sim\mathcal U(\pi/6,\pi)$. Spatial dependence across meta-atoms is induced by these shared path gains and array responses, without imposing an additional exponential spatial-correlation matrix. In particular, conditioned on the path angles,
\[
\mathbb E[\bm g_n\bm g_n^H\mid\{\varphi_{n,p},\vartheta_{n,p}\}]
=\frac{\bar g_{L,n}}{P_{\mathrm{path}}}
\sum_{p=1}^{P_{\mathrm{path}}}\bm a_{n,p}\bm a_{n,p}^H,
\]
where $\bm a_{n,p}=\bm a(\varphi_{n,p},\vartheta_{n,p})$. Electromagnetic mutual coupling within a layer is not explicitly modeled.
}
\subsubsection{Large-Scale Fading}
The link distance between the $n$-th user and the SIM is modeled as
\begin{align}
    d_n =& {\color{blue}\sqrt{r_n^2 + (h - D-d_{\mathrm{layer}})^2}}, \nonumber\\
    &\qquad r_n \sim \mathcal{U}(d_{r,\min}, d_{r,\max}),\, n\in\mathcal{N},
\end{align}
where $h$ is the deployment height of the BS and SIM, $D$ denotes the thickness of the SIM, {\color{blue}and $D+d_{\mathrm{layer}}$ includes the BS-to-first-layer separation in the distance offset used for the simulations,}
and $r_n$ is the horizontal distance between the SIM and the $n$-th user, which is uniformly distributed over $[d_{r,\min}, d_{r,\max}]$.
The corresponding large-scale fading coefficient is given by
\begin{equation}
    \bar{g}_{L,n} = 10^{C_0/10} \, d_n^{-\mu},
\end{equation}
where $C_0$ denotes the reference path loss (in dB) at unit distance, and $\mu$ is the path loss exponent. These coefficients reflect user-specific attenuation levels due to geometric spreading and propagation loss.

\subsection{Transmission Model}
Thus, the overall equivalent downlink channel between the BS antennas and $N$ UEs becomes
\begin{equation}
    \mathbf{H}_0\mathbf{\Phi}_1\mathbf{H}_1\cdots\mathbf{H}_{L-1}\mathbf{\Phi}_L\mathbf{G} = \mathbf{BG} \in \mathbb{C}^{N \times N},
\end{equation}
where $\mathbf{H}_l\quad (l\in\mathcal{L})$ are all known by the BS, $\mathbf{\Phi}_l\quad (l\in\mathcal{L})$ are controllable parameters, and $\mathbf{G}$ is an unknown channel.
Denoting the transmitted signal as $\mathbf{X}\in\mathcal{C}^{N\times T}$ over $T$ time slots, the received signal at all UEs becomes
\begin{equation}
    \mathbf{Y}^\top = \mathbf{G}^\top \mathbf{B}^\top\mathbf{X}+\mathbf{N},
\end{equation}
where $\mathbf{N}$ is the noise. The $i$-th row of $\mathbf{Y}$ corresponds to UE $i$, and the $j$-th column corresponds to time slot $j$, respectively.
{\color{blue}For data transmission, the desired signal power at UE $i$ is $p_i|\bm g_i^\top\mathbf B^\top\bm e_i|^2$, while the residual interference is $\sum_{j\ne i}p_j|\bm g_i^\top\mathbf B^\top\bm e_j|^2$. Treating interference as noise gives the corresponding SINR and achievable rate $\log_2(1+\mathrm{SINR}_i)$. Under exact ZF, this interference term vanishes and the expression reduces to the SNR-based objective below.}
\subsection{Problem Formulation}
Considering the ZF requirements, the optimal beamforming problem can be formulated as
\begin{subequations} \label{eq:opt}
\begin{align}
\argmax_{\{p_i\}_1^N,\{\bm{\Phi}_l\}_1^L} \quad & 
\sum_{i\in\mathcal{N}} \log_2 \!\left( 1 + \sigma^{-2} p_i 
|\bm{g}_{i}^\top \mathbf{B}^\top\bm{e}_i|^2\right), \label{eq:opt_obj} \\
\text{s.t.} \quad 
& \sum_{i=1}^N p_i \leq P_{\max}, \label{eq:opt_power} \\
& |\phi_{l,m}|\leq 1, \quad\forall l\in\mathcal{L},m\in\mathcal{M},\label{eq:opt_phi} \\
& \mathbf{BG}\; \textrm{is diagonal, (ZF requirement)} \label{eq:opt_zf}
\end{align}
\end{subequations}
where $\bm{g}_i$ denotes the $i$-th column of the SIM-UE channel $\mathbf{G}$ corresponding to UE $i$, and $\bm{e}_i$ denotes an $N$-dimensional one-hot vector where the $i$-th element is 1 and the others are 0s. 
{\color{blue}We emphasize that the ZF requirement is introduced primarily to provide a structured interference-suppression objective for the proposed gradient-based framework, rather than to claim that strict ZF is sum-rate optimal. Direct optimization of the sum-rate objective leads to a highly coupled non-convex landscape in which useful-signal enhancement and inter-user interference suppression are intertwined, which can result in unstable gradient updates. In contrast, the ZF structure explicitly separates the interference-suppression objective and allows it to be incorporated through a tractable penalty term in the subsequent gradient formulation.}
We also note that the last ZF requirement \eqref{eq:opt_zf} cannot be directly calculated due to unknown $\mathbf{G}$, which is otherwise indirectly addressed by loss design as shown below.
\section{Proposed Method Based on UE-Side Zero-Forcing}
\begin{figure}[h]
  \centering
  \includegraphics[width=0.85\linewidth]{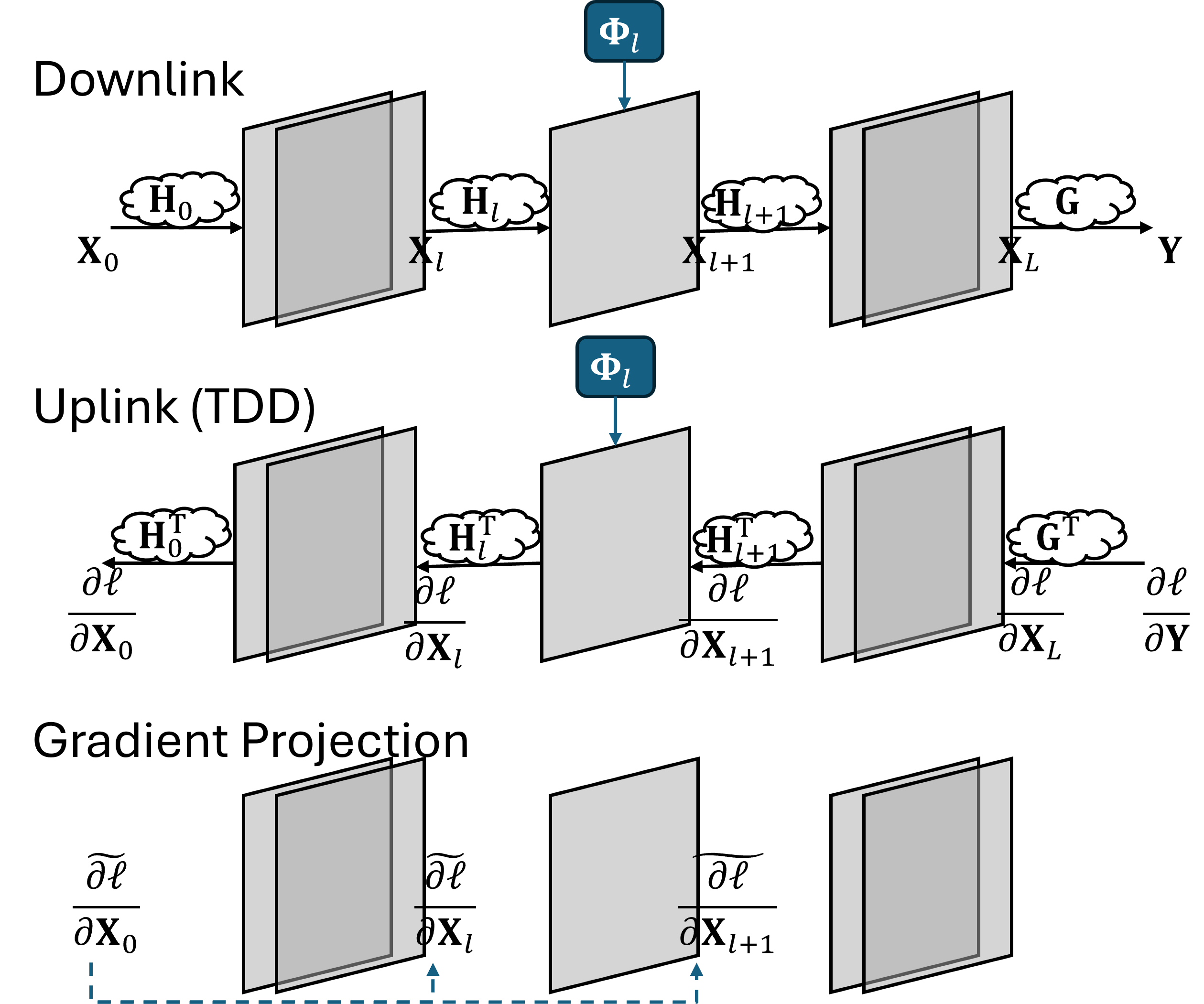}
  \caption{\textcolor{blue}{Workflow of the proposed P2GD method: downlink forward propagation through the SIM, gradient backpropagation over the reciprocal TDD uplink, and BS-side projected-gradient reconstruction for each SIM layer.}}
  \label{fig:relation}
\end{figure}
In this section, we delve into the proposed algorithm. Considering the channel reciprocity, we have two equivalent expressions of the ZF requirement. Specifically, we discuss the UE-side variation in this section, which follows the straightforward ZF procedure where the cross-UE downlink interference is minimized. The BS-side ZF variation, which eliminates the uplink interference among BS antennas, is later discussed in the next section. The main algorithm work flow is illustrated in Fig. \ref{fig:relation}. Specifically, the proposed system is optimized through trials instead of pre-obtained channel estimation. The UEs can then calculate the equivalent loss and initiate gradient backpropagation. As we cannot directly obtain the raw gradient results on the metasurfaces, we need to employ another separate pseudo-inverse mapping to imitate the real gradient values.

{\color{blue} Throughout the proposed procedure, the BS knows the SIM structure, the BS–SIM propagation matrices, and the transmitted pilot signals, while the SIM–UE channel and the intermediate signals inside the SIM are not directly observed. The UEs compute the local loss gradients from their received signals and transmit the corresponding feedback over the reciprocal uplink, after which the BS reconstructs a projected gradient using the known forward model and pseudo-inverse mapping.}
\subsection{Downlink Pilot Transmission}
To distinguish the signal transmitted to different UEs, we need to transmit a group of orthogonal pilots.
Specifically, we transmit a DFT matrix $\mathbf{F}:=\operatorname{DFT(N)}$ where
\begin{equation}
F_{m,n} = \exp\left(j2(m-1)(n-1)\pi/N\right)  
\end{equation}
{\color{blue}for $m,n\in\{1,\ldots,N\}$ in each transmission. We define $\mathbf P=\operatorname{diag}(p_1,\ldots,p_N)$ and $\mathbf P^{1/2}=\operatorname{diag}(\sqrt{p_1},\ldots,\sqrt{p_N})$.}
Assuming that the BS transmits $\mathbf{P}^{1/2}\mathbf{F}$ through the downstream,
for the simplification of the expression, we denote
\begin{equation}
\mathbf{X}^\top_l:=\mathbf{\Phi}_l\mathbf{H}^\top_{l-1}\cdots\mathbf{\Phi}_1\mathbf{H}^\top_0\mathbf{P}^{1/2}\mathbf{F}\label{Xdef}
\end{equation}
as the transmitted signal after the $l$-th layer of the SIM. We note that $\mathbf{X}_l$s cannot be measured directly due to hardware restriction. However, they can be predicted by the BS.

Thus, the UE-side received signal on UE $i$ becomes
\begin{equation}
\bm{y}_i^\top=\mathbf{g}_i^\top\mathbf{B}^{\top}\mathbf{P}^{1/2}\mathbf{F} +\mathbf{n}^\top_i,\;i\in\mathcal{N}\label{down:y}
\end{equation}
where $\mathbf{n}_i\sim\mathcal{CN}(\bm{0}, \sigma^2\mathbf{I})$ is the channel noise.
Further considering the ZF requirement, we know that {\color{blue} under the perfect ZF condition in \eqref{eq:opt_zf}, the effective channel $\mathbf G^{\top}\mathbf B^{\top}$ is diagonal. Hence, for UE $i$, the contributions from all streams $j\neq i$ vanish, and the $\tau$-th received pilot symbol becomes}
\begin{equation}
{\color{blue}y_{i,\tau}=\sqrt{p_i}\,\bm g_i^{\top}\mathbf B^{\top}\bm e_i F_{i,\tau}+n_{i,\tau}.}
\end{equation}
Since {\color{blue}$|F_{i,\tau}|^2=1$} for the adopted DFT pilots and
{\color{blue}$\mathbb{E}[|n_{i,\tau}|^2]=\sigma^2$}, we have
\begin{equation}{\color{blue}
\mathbb{E}\!\left[\bm y_i^{H}\bm y_i\right]=
N p_i \left|\bm g_i^{\top}\mathbf B^{\top}\bm e_i\right|^2+N\sigma^2,}
\end{equation}
which yields
\begin{equation}
{\color{blue}\mathbb{E}\{\bm{y}_i^{\mr H}\bm{y}_i\}}=N\sigma^2\left( 1 + \sigma^{-2} p_i 
|\bm{g}_{i}^\top \mathbf{B}^\top\bm{e}_i|^2\right).
\end{equation}
{\color{blue}In practical optimization, the instantaneous received pilot energy \(\bm y_i^H\bm y_i\) is directly observed and used in the local loss. Equations above only establish its expected correspondence to the original rate expression.}
\begin{equation}
\argmax_{\{p_i\}_1^N,\{\bm{\Phi}_l\}_1^L} \quad 
\sum_{i\in\mathcal{N}} \log \bm{y}_i^{\mr H}\bm{y}_i,\quad \mr{s.t.} \eqref{eq:opt_power}, \eqref{eq:opt_phi}, \eqref{eq:opt_zf}.\label{eq_opt2}
\end{equation}

\subsection{UE-Side Zero Forcing}
We first discuss realizing zero-forcing from the UE's side. 
Thus, the zero-forcing principle can be transformed to
\begin{equation}
\sum_{\substack{i,j \in\mathcal{N} \\ i \neq j}}\bm{g}_i^\top\mathbf{B}^{\top}\mathbf{P}^{1/2}\mathbf{F}\bm{f}_j=0. \label{UE-ZF}
\end{equation}
Substituting \eqref{down:y} into \eqref{UE-ZF} and ignoring the noise item, we can define the loss function
\begin{equation}\color{blue}
\ell_\mr{ZF}^\mr{UE}:=\sum_{i\in\mathcal{N}}\log\left(\epsilon_{\mathrm{ZF}}+\sum_{j\in\mathcal{N}, j\neq i}|\bm{y}_i^\top\bm{f}_j|^2\right),\label{loss_UE_ZF}
\end{equation}
{\color{blue}where $\epsilon_{\mathrm{ZF}}>0$ is a small regularization constant introduced to avoid numerical singularity when the residual interference approaches zero.}
It is incorporated into the original objective through the following penalized problem:
\begin{equation}\label{eq:opt3}
\argmax_{\{p_i\}_1^N,\{\bm{\Phi}_l\}_1^L} \quad 
\sum_{i\in\mathcal{N}} \log\bm{y}_i^{\mr H}\bm{y}_i-\lambda\ell_\mr{ZF}^\mr{UE}, \quad
\text{s.t.} \quad 
\eqref{eq:opt_power}, \eqref{eq:opt_phi},
\end{equation}
where $\lambda>0$ is the penalty coefficient. {\color{blue}Here, the logarithmic form is retained to keep the interference-suppression term on a comparable scale with the rate-related objective, while the regularization has negligible influence when the residual interference is appreciably larger than $\epsilon_{\mathrm{ZF}}$. Accordingly, the ZF penalty in \eqref{loss_UE_ZF} only encourages interference suppression rather than enforcing an exact interference-free solution.}

\subsection{Uplink Transmission}
Denote $\bm{v}_i^\top$ as the uplink transmitted signal from each UE over time, respectively. Also denote $\mathbf{U}_l$ the uplink signal received by the SIM layer $l$.
{\color{blue}Under the ideal TDD reciprocity assumption stated in Section I-A,} we have \cite{JSAC}
\begin{align}
\mathbf{U}^\top_L=&\sum_{i\in\mathcal{N}} \bm{g}_i\bm{v}_i^\top,&\label{back1}\\
\mathbf{U}^\top_l=&\mathbf{H}_l\mathbf{\Theta}_{l+1}\cdots\mathbf{H}_{L-1}\mathbf{\Theta}_L\mathbf{U}^\top_L,&l\in\mathcal{L}\backslash\{L\},\label{back2}\\
\mathbf{U}^\top_0=&\mathbf{H}_0\mathbf{\Theta}_1\mathbf{U}^\top_1+\mathbf{Z},&\label{back3}
\end{align}
where ${\mathbf Z}\sim\mathcal{CN}(\bm{0}, \sigma^2\mathbf{I})$ denotes the uplink noise, and $\mathbf{\Theta}_l$ denotes the uplink SIM coefficient matrix. {\color{blue}It is set equal to the downlink matrix $\mathbf{\Phi}_l$ in the proposed reciprocal procedure; the two symbols distinguish transmission direction, not independent physical controls.} We note that in practical transmissions, we can only obtain $\mathbf{U}_0$ on the BS side. Moreover, as proven in \cite{JSAC}, noisy gradients do not degrade the gradient method significantly. Thus, in the following algorithm design, we can omit the noise item ${\mathbf Z}$ for simplicity, since introducing the background noise in gradient calculation does not significantly deteriorate the performance.

\subsection{Gradient Backpropagation}
In this section, we examine the gradient of the target in \eqref{eq:opt3}. Firstly, denoting the optimization target
\begin{equation}
\ell_1:=\sum_{i\in\mathcal{N}} \log\bm{y}_i^{\mr H}\bm{y}_i-\lambda\ell_\mr{ZF}^\mr{UE},
\end{equation}
we know
\begin{equation}
\begin{aligned}
\frac{\partial \ell_1}{\partial \bm{y}^*_i}=&(\bm{y}_i^{\mr H}\bm{y}_i)^{-1}\bm{y}_i-\\
&\lambda\left({\color{blue}\epsilon_{\mathrm{ZF}}+}\sum^N_{j=1, j\neq i}|\bm{f}_j^{\mr H}\bm{y}_i^*|^2\right)^{-1}\sum^N_{j=1, j\neq i}(\bm{f}_j^{\mr H}\bm{y}_i^*)\bm{f}_j. \label{bp:y}
\end{aligned}
\end{equation}
We note that, given the received signal, \eqref{bp:y} can be independently computed for each UE.

Following the chain rule of backpropagation, we have
\begin{eqnarray}
\frac{\partial \ell_1}{\partial \mathbf{X}^*_L}=&\sum_{i\in\mathcal{N}}{\color{blue}\frac{\partial \ell_1}{\partial \bm{y_i}^*}}\bm{g}_i^\mr{H}, &\label{BP1}\\
\frac{\partial \ell_1}{\partial \mathbf{X}^*_{l-1}}=&\frac{\partial \ell_1}{\partial \mathbf{X}^*_l}\mathbf{\Phi}^\mr{H}_l\mathbf{H}^\mr{H}_{l-1}, &l\in\mathcal{L}, \label{BP2}\\
\frac{\partial \ell_1}{\partial \mathbf{\Phi}^*_l}=&\mathbf{H}^{\mr H}_{l-1}\mathbf{X}^\mr{H}_{l-1}\frac{\partial \ell_1}{\partial \mathbf{X}^*_l}, & l\in\mathcal{L}, \label{BP3}\\
\frac{\partial \ell_1}{\partial \mathbf{P}}=&\frac{1}{2}\mathbf{P}^{-1/2}\mathbf{F}^*\frac{\partial \ell_1}{\partial \mathbf{X}^*_0}. &\label{BP4}
\end{eqnarray}

We first compare \eqref{BP1}-\eqref{BP3} with \eqref{back1} and \eqref{back2}. We know that if we set
\begin{eqnarray}
\bm{v}_i=&\alpha\frac{\partial \ell_1}{\partial \bm{y_i}},&i\in\mathcal{N},\label{param1}\\
\mathbf{\Theta}_l=&\mathbf{\Phi}_l, &l\in\mathcal{L},\label{param2}
\end{eqnarray}
where $\alpha$ is a positive number to adjust the uplink transmission power to a reasonable range. {\color{blue}For the real-valued objective, $\partial\ell_1/\partial\bm y_i=(\partial\ell_1/\partial\bm y_i^*)^*$; hence the feedback uses the conjugate of the locally computed Wirtinger gradient.} We have
\begin{equation}
\frac{\partial \ell_1}{\partial \mathbf{X}^*_l}=\alpha^{-1}\mathbf{U}_l^*,l\in\mathcal{L}\cup\{0\},
\end{equation}
where only $\mathbf{U}_0$ can be obtained by the BS, and the noise item in \eqref{back3} is omitted.

We denote
\begin{equation}
\mathbf{B}_l:=\mathbf{H}_0\mathbf{\Phi}_1\cdots\mathbf{H}_{l-1}\mathbf{\Phi}_l.
\end{equation}
Using the pseudoinverse of the known forward mapping, we reconstruct the projected intermediate signal as
\begin{eqnarray}
&\widetilde{\mathbf{U}}_0=&\mathbf{U}_0,\label{final1}\\
&\widetilde{\mathbf{U}}_l=&\mathbf{U}_0\left(\mathbf{B}_l^\top\right)^\dagger,\label{final2}\\
&\widetilde{\mathbf{\Psi}}_l=&\mathbf{H}_{l-1}^{\mr H}\mathbf{X}^\mr{H}_{l-1}\widetilde{\mathbf{U}}_l^*,\label{final3}
\end{eqnarray}
which yields regarding \eqref{BP3}
\begin{eqnarray}
\widetilde{\mathbf{U}}^*_l=&\alpha\frac{\partial \ell_1}{\partial \mathbf{X}^*_l}\left(\mathbf{B}^\dagger_l\mathbf{B}_l\right)^\mr{H}
\label{pseudo-grad1}\\
\widetilde{\mathbf{\Psi}}_l=&\alpha\frac{\partial \ell_1}{\partial \mathbf{\Phi}^*_l}\left(\mathbf{B}^\dagger_l\mathbf{B}_l\right)^\mr{H}.\label{pseudo-grad2}
\end{eqnarray}
We note that \eqref{pseudo-grad1} and \eqref{pseudo-grad2} become computable, and with them, we obtain an approximation of a projected gradient.
\subsection{General Algorithm}
\begin{algorithm}[t]
\caption{UE-P2GD Algorithm}
\label{alg::1}
\begin{algorithmic}[1]
\STATE \textbf{Input:} Maximum transmission power $P_\mr{max}$ for BS, \ $P_\mr{UE}$ for UEs, penalty coefficient $\lambda$, hyperparameter $\alpha$, {\color{blue}learning rates $\eta_P$ and $\eta_\Phi$}
\STATE \textbf{Initialization:} BS randomly sample $\mathbf{\Phi}_l$ such that $\phi_{l,i}=e^{j\psi_{l,i}}$, all $\psi$s are independently drawn from $\mathcal{U}(0,2\pi)$. Initial transmission power $p_i=P_\mr{max}/N$
\FOR{{\color{blue}$t=1,\ldots,T_{\max}$}}
\STATE BS sets SIM parameters to $\{\mathbf{\Phi}_l\}_{l=1}^L$
\STATE BS downlink transmit $\mathbf{P}^{1/2}\mathbf{F}$ in $N$ time slots
\FOR{each UE in parallel}
\STATE UE $i$ receives $\bm{y}_i^\top$ as in \eqref{down:y}
\STATE {\color{blue}Calculate gradient $\bm{d}_i\leftarrow\partial\ell_1/\partial\bm{y}^*_i$} as \eqref{bp:y}
\FOR{{\color{blue}any $\tau$ such that $\alpha|d_{i,\tau}|>P_\mr{UE}$}}
\STATE {\color{blue}Clamping: $d_{i,\tau}\leftarrow \frac{P_\mr{UE}d_{i,\tau}}{\alpha|d_{i,\tau}|}$}
\ENDFOR
\STATE {\color{blue}Uplink transmit $\{\alpha d^*_{i,\tau}F_{i,\tau}\}_{\tau=1}^N$} in $N$ time slots
\ENDFOR
\STATE BS receives $\mathbf{U}_0$ as in \eqref{back1}-\eqref{back3}
\STATE Gradient of power $g_{p_i}\leftarrow\frac{\bm{f}_i^\mr{H}\bm{u}_{0,i}^*}{2\alpha\sqrt{p_i}}$ as \eqref{BP4}
\STATE Corresponding to \eqref{final1}-\eqref{final3}, $\mathbf{B}\leftarrow \mathbf{I}_N$
\STATE $\mathbf{X}_0\leftarrow\mathbf{F}\mathbf{P}^{1/2}, \mathbf{X}\leftarrow\mathbf{X}_0$
\FOR{$l=1,\cdots,L$}
\STATE $\mathbf{B}\leftarrow\mathbf{B}\mathbf{H}_{l-1}\mathbf{\Phi}_l$, $\mathbf{U}\leftarrow\mathbf{U}_0\left(\mathbf{B}^\top\right)^\dagger$
\STATE $\Psi_{l,i}\leftarrow\mathbf{H}_{l-1}^{\mr H}\bm{x}_i^\mr{H}\bm{u}_i^*, \mathbf{\Psi}_l:=\operatorname{diag}(\Psi_{l,1},\cdots,\Psi_{l,M})$
\STATE $\mathbf{X}\leftarrow\mathbf{X}_0\mathbf{B}$
\STATE {\color{blue}$\mathbf{\Phi}_l\leftarrow\mathbf{\Phi}_l+\eta_\Phi\alpha^{-1}\mathbf{\Psi}_l$}
\FOR{any $i$ such that $|\phi_{l,i}|>1$}
\STATE Apply power restriction: $\phi_{l,i}\leftarrow\phi_{l,i}/|\phi_{l,i}|$
\ENDFOR
\ENDFOR
\STATE {\color{blue}$p_i\leftarrow \max\{p_i+\eta_P g_{p_i}, 0\}$} \textbf{for} $i=1,\cdots,N$
\STATE $p_i\leftarrow P_\mr{max}p_i/\sum_{i=1}^Np_i$ \textbf{for} $i=1,\cdots,N$
\ENDFOR
\end{algorithmic}
\end{algorithm}
Finally, we come to the overall gradient ascent algorithm as shown in Algorithm \ref{alg::1}. The gradient regarding the optimization target in problem \eqref{eq:opt3} can be obtained through \eqref{BP4} and \eqref{pseudo-grad2}. As the remaining restrictions of \eqref{eq:opt_phi} and \eqref{eq:opt_power} are simple convex bounds over the target variables, we can directly add them in the gradient ascent process, where we choose different learning rates for power $\mathbf{P}$ and reflection coefficients $\{\mathbf{\Phi}_l\}$s for numerical stability. We note that as the variables are diagonal, we only need to calculate the diagonal elements in \eqref{BP4} and \eqref{final3}. Also, we note that we apply a maximum uplink power restriction when $\bm{y}_i^H\bm{y}_i$ becomes too small in the algorithm.
\section{Proposed Method Based on BS-Side Zero-Forcing}
In this section, we try to apply the ZF conditions on the BS side.
We adopt the same downlink transmission model as \eqref{down:y}.
Instead of directly transforming the ZF requirement into a penalty item, we consider the received power of the desired signal directly. Thus, different from \eqref{eq_opt2}, the optimization problem can be written as
\begin{equation}
\begin{aligned}
\argmax_{\{p_i\}_1^N,\{\bm{\Phi}_l\}_1^L}& \quad 
\ell_\mr{UE}:=\sum_{i=1}^N \log \bm{y}_i^{\mr H}\bm{f}_i\bm{f}_i^\mr{H}\bm{y}_i,\\
\mr{s.t.}&\quad \eqref{eq:opt_power}, \eqref{eq:opt_phi}, \eqref{eq:opt_zf}.
\end{aligned}\label{eq_opt4}
\end{equation}

It is easy to know
\begin{equation}
\frac{\partial\ell_\mr{UE}}{\partial\bm{y}^*_i}=(\bm{y}_i^\mr{H}\bm{f}_i)^{-1}\bm{f}_i,\label{bp_y2}
\end{equation}
which is along the direction of $\bm{f}_i$.
Utilizing the channel reciprocity,{\color{blue} according to \eqref{param1} and \eqref{bp_y2}, the uplink feedback vector of UE $i$ is given by
\begin{equation}
\bm v_i
=
\alpha
(\bm y_i^\top\bm f_i^*)^{-1}\bm f_i^*,
\end{equation}
which is parallel to the conjugated $i$-th DFT pilot vector $\mathbf f_i^*$.
Ignoring the uplink noise, \eqref{back1}-\eqref{back3} can be equivalently written as
\begin{equation}
\mathbf U_0
=
\mathbf V\mathbf G^{\top}\mathbf B^{\top},
\end{equation}
where $\mathbf V=[\mathbf v_1,\ldots,\mathbf v_N]$.
Under the ZF condition in \eqref{eq:opt_zf}, $\mathbf G^{\top}\mathbf B^{\top}$ is diagonal. Therefore, denoting $\mathbf u_{0,i}$ as the $i$-th column of $\mathbf U_0$, we have $\mathbf u_{0,i}\propto\mathbf v_i\propto\mathbf f_i^*$.
Owing to the orthogonality of the DFT pilots, this yields}
\begin{equation}
\mathbb{E}\{\bm{u}^\mr{H}_{0,i}\bm{f}_j^*\}=0, \quad\forall i\neq j.
\end{equation}
Therefore, we can write the following penalty item:
\begin{equation}
\ell_\mr{ZF}^\mr{BS}:=\sum_{i=1}^N \log\left({\color{blue}\epsilon_{\mathrm{ZF}}+}\sum^N_{j=1, j\neq i} |\bm{u}^\top_{0,i}\bm{f}_j|^2\right).
\end{equation}
Therefore, we have the optimization target
\begin{equation}
\ell_2:=\ell_\mr{UE}-\lambda\ell_\mr{ZF}^\mr{BS}.
\end{equation}

Through gradient backpropagation, noting that $\mathbf{\Phi}_l=\mathbf{\Theta}_l$, we know that
\begin{eqnarray}
\frac{\partial{\ell_\mr{ZF}^\mr{BS}}}{\partial\bm{u}^*_{0,i}}=&\left({\color{blue}\epsilon_{\mathrm{ZF}}+}\sum^N_{j=1, j\neq i} |\bm{u}^\top_{0,i}\bm{f}_j|^2\right)^{-1}\nonumber\\&\times\sum^N_{j=1, j\neq i}(\bm{u}^\top_{0,i}\bm{f}_j)\bm{f}_j^*,\label{final2_1}\\
\frac{\partial{\ell_\mr{ZF}^\mr{BS}}}{\partial\mathbf{U}^*_l}=& \frac{\partial{\ell_\mr{ZF}^\mr{BS}}}{\partial\mathbf{U}^*_{l-1}}\mathbf{H}^*_{l-1}\mathbf{\Phi}_l^*,\\
\frac{\partial{\ell_\mr{ZF}^\mr{BS}}}{\partial\mathbf{\Phi}^*_l}=&\mathbf{H}^\mr{H}_{l-1}\left(\frac{\partial{\ell_\mr{ZF}^\mr{BS}}}{\partial\mathbf{U}^*_{l-1}}\right)^\top\mathbf{U}_l\label{final2_3}.
\end{eqnarray}
However, $\mathbf{U}_l$ cannot be directly observed due to hardware restrictions. Similarly, we can use the pseudo inverse to obtain a projected gradient following \eqref{final1} and \eqref{final2}. Considering \eqref{Xdef}, \eqref{final1}-\eqref{final3}, and \eqref{final2_1}-\eqref{final2_3}, we can finally use the following process, where all required variables are known by the BS.
\begin{eqnarray}
&\widetilde{\mathbf{X}}_0=&\alpha^{-1}\mathbf{P}^{1/2}\mathbf{F}-\lambda\left(\frac{\partial{\ell_\mr{ZF}^\mr{BS}}}{\partial\mathbf{U}^*_0}\right)^*,\label{final3_1}\\
&\widetilde{\mathbf{X}}_l=&\widetilde{\mathbf{X}}_{l-1}\mathbf{H}_{l-1}\mathbf{\Phi}_l,\\
&\widetilde{\mathbf{\Psi}}_l=&\mathbf{H}_{l-1}^{\mr H}\widetilde{\mathbf{X}}^\mr{H}_{l-1}\widetilde{\mathbf{U}}_l^*,\label{final3_3}
\end{eqnarray}
where all $\widetilde{\mathbf{U}}_l$s are defined in \eqref{final1} and \eqref{final2}. It is easy to verify that, ignoring the background noise,
\begin{equation}
\widetilde{\mathbf{\Psi}}_l=\frac{\partial \ell_2}{\partial \mathbf{\Phi}^*_l}\left(\mathbf{B}^\dagger_l\mathbf{B}_l\right)^\mr{H}.
\end{equation}
Thus, we have the overall projected gradient ascent algorithm as shown in Algorithm \ref{alg::2}.\begin{algorithm}[t]
\caption{BS-P2GD Algorithm}
\label{alg::2}
\begin{algorithmic}[1]
\STATE \textbf{Input:} Maximum transmission power $P_\mr{max}$ for BS, \ $P_\mr{UE}$ for UEs, penalty coefficient $\lambda$, hyperparameter $\alpha$, {\color{blue}learning rates $\eta_P$ and $\eta_\Phi$}
\STATE \textbf{Initialization:} BS randomly sample $\mathbf{\Phi}_l$ such that $\phi_{l,i}=e^{j\psi_{l,i}}$, all $\psi$s are independently drawn from $\mathcal{U}(0,2\pi)$. Initial transmission power $p_i=P_\mr{max}/N$
\FOR{{\color{blue}$t=1,\ldots,T_{\max}$}}
\STATE BS sets SIM parameters to $\{\mathbf{\Phi}_l\}_{l=1}^L$
\STATE BS downlink transmit $\mathbf{P}^{1/2}\mathbf{F}$ in $N$ time slots
\FOR{each UE \textbf{in parallel}}
\STATE UE $i$ receives $\bm{y}_i^\top$ as in \eqref{down:y}
\STATE {\color{blue}Calculate gradient $\bm{d}_i\leftarrow(\bm{y}_i^\mr{H}\bm{f}_i)^{-1}\bm{f}_i$}
\FOR{{\color{blue}any $\tau$ such that $\alpha|d_{i,\tau}|>P_\mr{UE}$}}
\STATE {\color{blue}Clamping: $d_{i,\tau}\leftarrow \frac{P_\mr{UE}d_{i,\tau}}{\alpha|d_{i,\tau}|}$}
\ENDFOR
\STATE {\color{blue}Uplink transmit $\{\alpha d^*_{i,\tau}F_{i,\tau}\}_{\tau=1}^N$} in $N$ time slots
\ENDFOR
\STATE BS receives $\mathbf{U}_0$ as in \eqref{back1}-\eqref{back3}
\STATE Gradient of power $g_{p_i}\leftarrow\Re\frac{\bm{f}_i^\mr{H}\bm{u}_{0,i}^*}{2\alpha\sqrt{p_i}}$ as \eqref{BP4}
\STATE Corresponding to \eqref{final1},\eqref{final2} \& \eqref{final3_1}-\eqref{final3_3}, $\mathbf{U}\leftarrow \mathbf{U}_0$
\STATE $\mathbf{X}\leftarrow\widetilde{\mathbf X}_0$ as defined in \eqref{final3_1}, $\mathbf{B}\leftarrow \mathbf{I}_N$
\FOR{$l=1,\cdots,L$}
\STATE $\mathbf{B}\leftarrow\mathbf{B}\mathbf{H}_{l-1}\mathbf{\Phi}_l$, $\mathbf{U}\leftarrow{\color{blue}\mathbf{U}_0}\left(\mathbf{B}^\top\right)^\dagger$
\STATE $\Psi_{l,i}\leftarrow\mathbf{H}_{l-1}^{\mr H}\bm{x}_i^\mr{H}\bm{u}_i^*, \mathbf{\Psi}_l:=\operatorname{diag}(\Psi_{l,1},\cdots,\Psi_{l,M})$
\STATE $\mathbf{X}\leftarrow\mathbf{X}_0\mathbf{B}$
\STATE {\color{blue}$\mathbf{\Phi}_l\leftarrow\mathbf{\Phi}_l+\eta_\Phi\mathbf{\Psi}_l$}
\FOR{any $i$ such that $|\phi_{l,i}|>1$}
\STATE Apply power restriction: $\phi_{l,i}\leftarrow\phi_{l,i}/|\phi_{l,i}|$
\ENDFOR
\ENDFOR
\STATE {\color{blue}$p_i\leftarrow \max\{p_i+\eta_P g_{p_i}, 0\}$} \textbf{for} $i=1,\cdots,N$
\STATE $p_i\leftarrow P_\mr{max}p_i/\sum_{i=1}^Np_i$ \textbf{for} $i=1,\cdots,N$
\ENDFOR
\end{algorithmic}
\end{algorithm}
\section{Theoretical Results}
\subsection{Projected Gradient Method}\label{sec:proj_grad}
{\color{blue}Here, we first analyze the convergence behavior of the projected-gradient mechanism under an idealized smooth convex setting, which provides a theoretical reference for the projected updates used in the proposed algorithms.} In this section, we use the Frobenius inner products and norms unless otherwise stated, i.e.,
\begin{equation}
\langle \mathbf{A}, \mathbf{B}\rangle:=\operatorname{tr}(\mathbf{A}^\mr{H}\mathbf{B}),\quad \|\mathbf{A}\|:=\sqrt{\langle\mathbf{A},\mathbf{A}\rangle}.
\end{equation}

We first consider a convex optimization problem with an arbitrary matrix $\mathbf{A}$
\begin{equation}
\min_\mathbf{X} \ell(\mathbf{B}), \quad \mathbf{B}:=\mathbf{AX},\label{prob:proof1}
\end{equation}
where $\ell$ is real-valued with some lower bound, convex, differentiable.
We denote $\partial_{\mathbf{M}}\ell:=\partial\ell/\partial\mathbf{M}$ for any $\mathbf{M}$ related to \eqref{prob:proof1}. 
We assume that $\ell(\mathbf{B})$ is $\beta$-smooth such that for some $\beta$,
\begin{equation}
\|\partial_{\mathbf{B}}\ell(\mathbf{B}_1) - \partial_{\mathbf{B}} \ell(\mathbf{B}_2)\| \leq \beta \|\mathbf{B}_1-\mathbf{B}_2\|, \quad \forall \mathbf{B}_1,\mathbf{B}_2 \in \mathbb{C}^{N\times M}.
\end{equation}

{\color{blue}The variable $\mathbf X$ in this subsection is an abstract optimization matrix, distinct from the transmitted and intermediate signals denoted by $\mathbf X$ and $\mathbf X_l$ in the system model.}

To solve the problem \eqref{prob:proof1}, we consider the following projected gradient descent method beginning from any $\mathbf{X}_0$:
\begin{equation}
\mathbf{X}_{t+1}=\mathbf{X}_t-\eta \left(\partial_{\mathbf{X}^*_t}\ell\right)\mathbf{Q}_t,\label{PGD}
\end{equation}
where we denote the projection matrix as
\begin{equation}
\mathbf{Q}_t:=\left(\mathbf{B}_t^\dagger\mathbf{B}_t\right)^\mr{H}
=\mathbf{X}_t^\mr{H}\mathbf{A}^\mr{H}\left(\mathbf{X}_t^\mr{H}\mathbf{A}^\mr{H}\right)^\dagger\label{pt}
\end{equation}

We can first verify the following lemma.
\begin{lemma}\label{lem:1}
\begin{equation}
\mathbf{Q}_t=\mathbf{Q}_t^2=\mathbf{Q}_t^\mr{H},
\end{equation}
where $\mathbf{Q}_t$ is defined in \eqref{pt}.
\end{lemma}
\begin{proof}
{\color{blue}Let the compact SVD be $\mathbf B_t=\mathbf U_r\mathbf\Sigma_r\mathbf V_r^H$, where the columns of $\mathbf V_r$ are orthonormal. Then $\mathbf B_t^\dagger\mathbf B_t=\mathbf V_r\mathbf V_r^H$, so $\mathbf Q_t=\mathbf V_r\mathbf V_r^H$ is Hermitian and $\mathbf Q_t^2=\mathbf V_r(\mathbf V_r^H\mathbf V_r)\mathbf V_r^H=\mathbf Q_t$. The zero-rank case gives $\mathbf Q_t=\mathbf0$ and satisfies the same identities.}
\end{proof}
\begin{remark}
Lemma \ref{lem:1} shows that the $\mathbf{Q}_t$ is an orthogonal projection matrix. Thus, $\mathbf{XQ}_t$ simply projects $\mathbf{X}$ to a low-dimensional linear space whose dimension is determined by the rank of $\mathbf{B}_t$.
\end{remark}

Considering the one-step gradient descent, we start from the following Wirtinger descent lemma in the complex domain, which is a trivial extension of the real version \cite{brandwood1983complex}.
\begin{lemma}\label{lem:descent}
(\textbf{Wirtinger descent lemma} \cite{brandwood1983complex}) For any function $f(\mathbf{X})$ which is differentiable and $\beta$-smooth, we have
\begin{equation}
f(\mathbf{X}_2)\leq f(\mathbf{X}_1)+2\Re\langle\partial_{\mathbf{X}^*}f(\mathbf{X}), \mathbf{X}_2-\mathbf{X}_1\rangle+\frac{\beta}{2}\|\mathbf{X}_2-\mathbf{X}_1\|^2.
\end{equation}
\end{lemma}
With Lemma \ref{lem:descent}, we can directly obtain the following theorem describing the convergence of the projected gradient method.
\begin{theorem}\label{the:1}
Let
\begin{eqnarray}
    0<\eta<\frac{4}{\beta\|\mathbf{A}\|^2},\label{PGD_eta}
\end{eqnarray}
the projected gradient descent algorithm in \eqref{PGD} always converges to some suboptimal points $\widehat{\mathbf{X}}$, where $\widehat{\mathbf{X}}$ satisfies
\begin{equation}
\left(\partial_{\widehat{\mathbf X}^*}\ell\right)\widehat{\mathbf X}^\mr{H}\mathbf{A}^\mr{H}\left(\widehat{\mathbf X}^\mr{H}\mathbf{A}^\mr{H}\right)^\dagger=\mathbf{0}.\label{PGD_target}
\end{equation}
Specifically, the optimal convergence rate satisfies
\begin{equation}
\min_{0\leq t\leq T}\left\|\left(\partial_{\widehat{\mathbf X}_t^*}\ell\right)\mathbf{Q}_t\right\|^2=\mathcal{O}(\beta\|\mathbf{A}\|^2/T)\label{convergence}
\end{equation}
when $\eta=\frac{2}{\beta\|\mathbf{A}\|^2}$.
\end{theorem}
\begin{proof}
From Lemma \ref{lem:1}, we know that for any matrix $\mathbf{X}$,
\begin{equation}
\begin{aligned}
\langle\mathbf{X},\mathbf{XQ}_t\rangle&=\langle\mathbf{X},\mathbf{XQ}^2_t\rangle=\langle\mathbf{XQ}_t^\mr{H},\mathbf{XQ}_t\rangle\\&=\langle\mathbf{XQ}_t,\mathbf{XQ}_t\rangle=\|\mathbf{XQ}_t\|^2    
\end{aligned}
\end{equation}

We can derive the error bound from Lemma \ref{lem:descent} that
\begin{equation}
\begin{aligned}
&\ell(\mathbf{B}_{t+1})-\ell(\mathbf{B}_t)\\
\leq&2\Re\langle\partial_{\mathbf{B}^*_t}\ell, \mathbf{A}(\mathbf{X}_{t+1}-\mathbf{X}_t)\rangle+\frac{\beta}{2}\|\mathbf{A}(\mathbf{X}_{t+1}-\mathbf{X}_t)\|^2\\
=&-2\eta\Re\langle\mathbf{A}^\mr{H}\left(\partial_{\mathbf{B}^*_t}\ell\right),\left(\partial_{\mathbf{X}^*_t}\ell\right)\mathbf{Q}_t\rangle+\frac{\beta\eta^2}{2}\|\mathbf{A}\left(\partial_{\mathbf{X}^*_t}\ell\right)\mathbf{Q}_t\|^2\\
=&-2\eta\Re\langle\left(\partial_{\mathbf{X}^*_t}\ell\right),\left(\partial_{\mathbf{X}^*_t}\ell\right)\mathbf{Q}_t\rangle+\frac{\beta\eta^2}{2}\|\mathbf{A}\left(\partial_{\mathbf{X}^*_t}\ell\right)\mathbf{Q}_t\|^2\\
=&-2\eta\|\left(\partial_{\mathbf{X}^*_t}\ell\right)\mathbf{Q}_t\|^2+\frac{\beta\eta^2}{2}\|\mathbf{A}\left(\partial_{\mathbf{X}^*_t}\ell\right)\mathbf{Q}_t\|^2\\
\leq&-2\eta\|\left(\partial_{\mathbf{X}^*_t}\ell\right)\mathbf{Q}_t\|^2+\frac{\beta\eta^2}{2}\|\mathbf{A}\|^2\|\left(\partial_{\mathbf{X}^*_t}\ell\right)\mathbf{Q}_t\|^2.\label{eq_proof}
\end{aligned}
\end{equation}
Thus, as $\ell$ has a lower bound and $\ell(\mathbf{B}_t)$ is decreasing, \eqref{PGD} converges to some points $\widehat{\mathbf{X}}$ that fulfills \eqref{PGD_target} with $\eta$ determined in \eqref{PGD_eta}.
Further, we know that
\begin{equation}
\sum_{t=0}^T \|\left(\partial_{\mathbf{X}^*_t}\ell\right)\mathbf{Q}_t\|^2 \leq \frac{2}{4\eta-\beta\eta^2\|\mathbf{A}\|^2}\left(\ell(\mathbf{AX}_0)-\ell(\mathbf{A}\widehat{\mathbf X})\right).
\end{equation}
Thus, when taking the optimal value $\eta=\frac{2}{\beta\|\mathbf{A}\|^2}$, we have
\begin{equation}
\frac{1}{T}\sum_{t=0}^T \|\left(\partial_{\mathbf{X}^*_t}\ell\right)\mathbf{Q}_t\|^2 \leq\mathcal{O}(\frac{\beta\|\mathbf{A}\|^2}{T}),
\end{equation}
which yields \eqref{convergence} and finishes the proof.
\end{proof}

{\color{blue}Theorem~\ref{the:1} characterizes the projected-gradient iteration in \eqref{PGD} under the smooth convex setting. Its purpose is to provide a theoretical reference for understanding the effect of the projection operation on the gradient update. However, the practical SIM beamforming problems considered in Algorithms~\ref{alg::1} and~\ref{alg::2}, are generally non-convex due to the cascaded SIM mapping, the logarithmic signal and interference terms, and the physical constraints on the SIM coefficients. Therefore, Theorem~\ref{the:1} only characterizes the behavior of the underlying projected-gradient mechanism under the stated idealized conditions.}
\begin{remark}
We can observe from \eqref{eq:propagation_coeff} that all $\mathbf{H}_l$s are almost of full rank. Moreover, we know that unless there is some $m$ such that $\phi_{l,m}=0$, $\mathbf{\Phi}_l$ is also of full rank. As $\phi_{l,m}$ can continuously take values in $\{x\in\mathcal{C}||x|\leq1\}$, we can say it is almost impossible that $\phi_{l,i}=0$ for any $l\in\mathcal{L}$ and $m\in\mathcal{M}$. Thus, the ranks of all $\mathbf{B}_l$s are almost the same as $\mathbf{H}_0$, which is determined by the number of BS antennas $N$. Thus, the performance of projected gradient descent is closer to the method in \cite{an2025stacked_TWC} when $N/M$ becomes relatively large. However, at the same time, the advantages in reducing transmission overhead also shrink when $N/M$ {\color{blue}increases,} since we need to transmit more pilots in each iteration. 

{\color{blue}Therefore, the proposed method is particularly suitable for regimes where the observable dimension $N$ is not excessively small relative to the number of SIM variables $M$. Increasing $N/M$ generally enriches the observable gradient subspace and reduces the potential projection loss, but simultaneously increases the pilot and signaling overhead. This reflects an intrinsic tradeoff between gradient fidelity and overhead reduction in the proposed framework.}
\end{remark}

\subsection{Diagonal Limitation of SIM Phase Matrices}
In this section, we aim to discuss the practical restrictions of SIMs. As shown above, we know that the proposed projected gradient algorithm can converge to a set of suboptimal $\overline{\mathbf \Phi}$. We then consider the specific structure determined by \eqref{phibar}. 
{\color{blue}Specifically, in practice, the aggregated SIM response
\begin{equation}
\overline{\mathbf \Phi}=f(\mathbf{\Phi}_1,\ldots,\mathbf{\Phi}_L):=\mathbf{\Phi}_1\mathbf{H}_1\cdots\mathbf{H}_{L-1}\mathbf{\Phi}_L\label{phibar}
\end{equation}
is restricted to an achievable set determined jointly by the diagonal structure and bounded entries of $\{\mathbf{\Phi}_l\}$, as well as by the ranks of the inter-layer propagation matrices $\{\mathbf{H}_l\}$. Therefore, the achievable set of $\overline{\mathbf{\Phi}}$ does not necessarily span the entire space $\mathbb{C}^{M\times M}$. To isolate the effect of the diagonal SIM update and simplify the analysis, we first relax this realizability constraint as follows.}
\begin{assumption} \label{ass}
{\color{blue}For the purpose of analysis, the aggregated matrix $\overline{\mathbf{\Phi}}$ defined in \eqref{phibar} is assumed to be freely variable in $\mathbb{C}^{M\times M}$, i.e., we neglect here the restriction that $\overline{\mathbf{\Phi}}$ must belong to the achievable set induced by the cascaded SIM structure.}
\end{assumption}
Under Assumption \ref{ass}, we have the following theorem. {\color{blue}It further characterizes the effect of the diagonal SIM structure under relaxed realizability assumption and the stated directional conditions. It provides sufficient analytical conditions for the structured projected update, rather than a global convergence guarantee for the original non-convex beamforming problem.}
\begin{theorem}\label{the:2}
Consider $\overline{\mathbf{\Phi}}$ as defined in \eqref{phibar}, denote
\begin{equation}
D(\overline{\mathbf{\Phi}}, \mathbf{\Delta}_{\overline{\Phi}},\eta):={\color{blue}\overline{\mathbf{\Phi}}-f(\mathbf{\Phi}_1-\eta\mathbf{\Delta}_{\Phi_1},\cdots,\mathbf{\Phi}_L-\eta\mathbf{\Delta}_{\Phi_L})},
\end{equation}
where
\begin{equation}
\mathbf{\Delta}_{\Phi_l}:=\mathcal{Q}_\mr{diag}\left(\mathbf{H}_{l-1}^\mr{H}\mathbf{\Phi}_{l-1}^\mr{H}\cdots\mathbf{H}_1^\mr{H}\mathbf{\Phi}_1^\mr{H}\mathbf{\Delta}_{\overline{\Phi}}\mathbf{\Phi}_L^\mr{H}\mathbf{H}_{L-1}^\mr{H}\cdots\mathbf{H}_l^\mr{H}\right),
\end{equation}
and $\mathcal{Q}_\mr{diag}\mathbf{X}$ is a projection operator that extracts the diagonal elements of $\mathbf{X}$ while retaining the shape, i.e.,
\begin{equation}
\left[\mathcal{Q}_\mr{diag}\mathbf{X}\right]_{i,j}=
\begin{cases}
X_{i,i}, & i=j,\\
0, & i\neq j.
\end{cases}
\end{equation}
We denote the angle between two matrices as \begin{equation}\angle(\mathbf{A},\mathbf{B}):=\arccos\frac{\Re\langle\mathbf{A},\mathbf{B}\rangle}{\|\mathbf{A}\|\|\mathbf{B}\|}.
\end{equation}

Thus, for any $0<\epsilon,\zeta<1$ and $0<\xi_1<1<\xi_2$, if we always have the following
\begin{equation}
\cos\angle\left( D(\overline{\mathbf{\Phi}}, \mathbf{\Delta}_{\overline{\Phi}},\eta), \mathbf{\Delta}_{\overline{\Phi}}\right)\geq\epsilon,\label{angle}
\end{equation}
\begin{equation}
(1-\epsilon^2)\left\|\partial_{\overline{\mathbf{\Phi}}^*_t}\ell\right\|^2 \leq \zeta\left\|\left(\partial_{\overline{\mathbf{\Phi}}^*_t}\ell\right) \mathbf{Q}_t\right\|^2,
\end{equation}
\begin{equation}
\xi_1\eta\|\mathbf{\Delta}_{\overline{\Phi}}\|\leq\|D(\overline{\mathbf{\Phi}}, \mathbf{\Delta}_{\overline{\Phi}},\eta)\|\leq\xi_2\eta\|\mathbf{\Delta}_{\overline{\Phi}}\|,
\end{equation}
\begin{equation}
0<\eta<\frac{4\epsilon-4\sqrt{\epsilon^2-1+\zeta}}{\beta\xi_2\|\mathbf{A}\|^2}\label{PGD2_eta},
\end{equation}
for some $\eta$, {\color{blue}the following method has a vanishing projected-gradient residual as in \eqref{PGD_target}, provided that the stated conditions hold throughout the iteration.}
\begin{equation}
\mathbf{X}_l=\mathbf{X}_l-\eta \mathcal{Q}_\mr{diag}\left(\left(\partial_{\mathbf{X}^*_l}\ell\right)\left(\mathbf{B}_l^\dagger\mathbf{B}_l\right)^{\mr H}\right).\label{PGD2}
\end{equation}
\end{theorem}
\begin{proof}
Following Lemma \ref{lem:descent}, we know
\begin{equation}
\begin{aligned}
&\ell(\mathbf{B}_{t+1})-\ell(\mathbf{B}_t)\\
\leq&\!-\!2\Re\langle\partial_{\overline{\mathbf{\Phi}}^*_t}\ell, {\color{blue}(\overline{\mathbf{\Phi}}_t-\overline{\mathbf{\Phi}}_{t+1})}\rangle+\frac{\beta}{2}\|\mathbf{A}\|^2\|(\overline{\mathbf{\Phi}}_{t+1}-\overline{\mathbf{\Phi}}_t)\|^2\\
=&\!-\!2\Re\langle\partial_{\overline{\mathbf{\Phi}}^*_t}\ell, D(\overline{\mathbf{\Phi}}_t, (\partial_{\overline{\mathbf{\Phi}}^*_t}\ell)\mathbf{Q}_t,\eta)\rangle\\
&+\frac{\beta}{2}\|\mathbf{A}\|^2\|D(\overline{\mathbf{\Phi}}_t, (\partial_{\overline{\mathbf{\Phi}}^*_t}\ell)\mathbf{Q}_t,\eta)\|^2.
\end{aligned}
\end{equation}

We note that since $\left\langle\partial_{\overline{\mathbf{\Phi}}^*_t}\ell, \left(\partial_{\overline{\mathbf{\Phi}}^*_t}\ell\right) \mathbf{Q}_t\right\rangle=\left\|\left(\partial_{\overline{\mathbf{\Phi}}^*_t}\ell\right) \mathbf{Q}_t\right\|^2$ always holds, we know 
\begin{equation}
   \angle\left(\partial_{\overline{\mathbf{\Phi}}^*_t}\ell, \left(\partial_{\overline{\mathbf{\Phi}}^*_t}\ell\right) \mathbf{Q}_t\right)=\arccos\frac{\|\left(\partial_{\overline{\mathbf{\Phi}}^*_t}\ell\right) \mathbf{Q}_t\|}{\|\partial_{\overline{\mathbf{\Phi}}^*_t}\ell\|}. 
\end{equation}
Thus, from \eqref{angle}, we know
\begin{equation}
\begin{aligned}
&\frac{\Re\langle\partial_{\overline{\mathbf{\Phi}}^*_t}\ell, D(\overline{\mathbf{\Phi}}_t, (\partial_{\overline{\mathbf{\Phi}}^*_t}\ell)\mathbf{Q}_t,\eta)\rangle}{\|\partial_{\overline{\mathbf{\Phi}}^*_t}\ell\|\|D(\overline{\mathbf{\Phi}}_t, (\partial_{\overline{\mathbf{\Phi}}^*_t}\ell)\mathbf{Q}_t,\eta)\|}\\
=&\cos\angle\left(\partial_{\overline{\mathbf{\Phi}}^*_t}\ell, D(\overline{\mathbf{\Phi}}_t, (\partial_{\overline{\mathbf{\Phi}}^*_t}\ell)\mathbf{Q}_t,\eta)\right)\\
\geq&\cos\left(\angle\left(\partial_{\overline{\mathbf{\Phi}}^*_t}\ell, \left(\partial_{\overline{\mathbf{\Phi}}^*_t}\ell\right) \mathbf{Q}_t\right)+\arccos\epsilon\right)\\
=&\frac{\epsilon\left\|\left(\partial_{\overline{\mathbf{\Phi}}^*_t}\ell\right) \mathbf{Q}_t\right\|\!-\!\sqrt{(1-\epsilon^2)\left(\|\partial_{\overline{\mathbf{\Phi}}^*_t}\ell\|^2\!-\!\left\|\left(\partial_{\overline{\mathbf{\Phi}}^*_t}\ell\right) \mathbf{Q}_t\right\|^2\right)}}{\|\partial_{\overline{\mathbf{\Phi}}^*_t}\ell\|}\\
\geq&\left(\epsilon-\sqrt{\epsilon^2-1+\zeta}\right)\frac{\left\|\left(\partial_{\overline{\mathbf{\Phi}}^*_t}\ell\right) \mathbf{Q}_t\right\|}{\|\partial_{\overline{\mathbf{\Phi}}^*_t}\ell\|}.
\end{aligned}
\end{equation}
Thus,
\begin{equation}
\begin{aligned}
&\ell(\mathbf{B}_{t+1})-\ell(\mathbf{B}_t)\\
\leq&{\color{blue}-2C_1}\left\|\left(\partial_{\overline{\mathbf{\Phi}}^*_t}\ell\right) \mathbf{Q}_t\right\|\|D(\overline{\mathbf{\Phi}}_t, (\partial_{\overline{\mathbf{\Phi}}^*_t}\ell)\mathbf{Q}_t,\eta)\|\\&+\frac{\beta\|\mathbf{A}\|^2}{2}\|D(\overline{\mathbf{\Phi}}_t, (\partial_{\overline{\mathbf{\Phi}}^*_t}\ell)\mathbf{Q}_t,\eta)\|^2\\
\leq&\left(-2C_1\!+\!\frac{\beta\xi_2\eta\|\mathbf{A}\|^2}{2}\right)\left\|\left(\partial_{\overline{\mathbf{\Phi}}^*_t}\ell\right)\!\! \mathbf{Q}_t\right\|\!\|D(\overline{\mathbf{\Phi}}_t, (\partial_{\overline{\mathbf{\Phi}}^*_t}\ell)\mathbf{Q}_t,\eta)\|,
\end{aligned}
\end{equation}
where $C_1:=\epsilon-\sqrt{\epsilon^2-1+\zeta}$. From \eqref{PGD2_eta}, we further know $C_2:=-2C_1+\frac{\beta\xi_2\eta\|\mathbf{A}\|^2}{2}<0$
\begin{equation}
\ell(\mathbf{B}_{t+1})-\ell(\mathbf{B}_t)
\leq\xi_1\eta C_2\left\|\left(\partial_{\overline{\mathbf{\Phi}}^*_t}\ell\right) \mathbf{Q}_t\right\|^2,
\end{equation}
{\color{blue}Since $\ell$ is bounded below and $\xi_1\eta C_2<0$, summing this inequality bounds the sum of squared projected-gradient residuals. Hence these residuals tend to zero, using the same telescoping argument as in the proof of Theorem~\ref{the:1}. This establishes the projected-stationarity conclusion under the stated directional conditions.}
\end{proof}

{\color{blue}The following remarks clarify the scope of the assumptions in Theorem~\ref{the:2}.}
\begin{remark}
{\color{blue}As $L$ increases, the achievable set of the cascaded SIM response generally becomes richer, making the relaxation in Assumption~\ref{ass} less restrictive. When the actual achievable set occupies only a subspace of the ambient matrix space, the same analysis can be interpreted within that reachable subspace.}
\end{remark}
\begin{remark}
{\color{blue}Conditions \eqref{angle}--\eqref{PGD2_eta} require alignment and magnitude control of the aggregate update. A small step size alone does not establish these properties for every cascaded SIM configuration. Theorem~\ref{the:2} is therefore conditional on these properties holding along the considered trajectory.}
\end{remark}
\subsection{Influence of Noise}
In the proposed method, we omitted the noise introduced in the backward transmission. Such noise is unavoidable in wireless systems. However, it is not always harmful. As shown in Theorem 1 of \cite{JSAC}, adding noise after each step does not affect the convergence of the gradient descent algorithm in convex problems while reducing the convergence rate by a constant multiple determined by the noise level.

Moreover, according to the saddle-escape theory \cite{ge2015escaping, du2017gradient}, introducing small random perturbations enables the iteration to escape such regions along directions of negative curvature and continue descending.
{\color{blue}Since $\ell_1$ and $\ell_2$ are non-convex with respect to the SIM coefficients, saddle points cannot be excluded in general. Moreover,} the projection matrix $\mathbf{Q}_t$ may introduce spurious saddle-like stationary regions near the constraint boundary, where the projected gradient vanishes but the true gradient does not. However, such points are non-stationary in the original space.
Consequently, the perturbed projected gradient method can reduce the possibility of falling into these artificial saddle points{\color{blue}, although convergence to a global optimum is not guaranteed}.

\subsection{Transmission and Computational Complexity}
{\color{blue}
We first consider the transmission overhead of the proposed methods. Let $T$ denote the number of P2GD iterations. In each iteration, both UE-P2GD and BS-P2GD require $N$ downlink MIMO symbols for orthogonal pilot transmission and $N$ uplink MIMO symbols for over-the-air gradient feedback. Therefore, the signaling overhead over a complete optimization episode is
\begin{equation}
\mathcal{O}_{\mathrm{DL}}^{\mathrm{P2GD}} = NT,\qquad
\mathcal{O}_{\mathrm{UL}}^{\mathrm{P2GD}} = NT.
\end{equation}
The downlink and uplink overheads are reported separately since the two directions may occupy different radio resources in a practical TDD system.

For comparison, conventional CSI-based beamforming requires explicit estimation of the SIM-UE channel $\mathbf{G}\in\mathbb{C}^{M\times N}$ before beamforming optimization. The channel contains $MN$ unknown complex coefficients. A MIMO pilot symbol observed over the $N$ BS antenna dimensions can provide at most $N$ independent complex observations. Hence, even under an ideal noise-free and fully identifiable measurement model, the number of MIMO pilot symbols required for explicit channel estimation satisfies
\begin{equation}
T_{\mathrm{CE}}N \geq MN,\qquad T_{\mathrm{CE}}\geq M.
\end{equation}
Therefore, $M$ MIMO pilot symbols are used as an ideal lower bound for explicit CSI acquisition in the following comparison. Practical channel estimation generally requires additional pilot observations in the presence of noise and model imperfections.

Besides air-interface signaling, each P2GD iteration may also incur latency associated with SIM reconfiguration, control signaling, synchronization, and UE-side gradient processing. Let $T_{\mathrm{DL}}$ and $T_{\mathrm{UL}}$ denote the duration of one downlink and uplink MIMO symbol, respectively, and let $T_{\mathrm{reconf}}$, $T_{\mathrm{ctrl}}$, $T_{\mathrm{sync}}$, and $T_{\mathrm{grad}}$ denote the corresponding per-iteration implementation delays. The end-to-end latency of a complete P2GD optimization episode can be characterized as
\begin{equation}
\tau_{\mathrm{P2GD}}
=
T\left(
NT_{\mathrm{DL}}
+
NT_{\mathrm{UL}}
+
T_{\mathrm{reconf}}
+
T_{\mathrm{ctrl}}
+
T_{\mathrm{sync}}
+
T_{\mathrm{grad}}
\right).
\end{equation}

For a channel coherence interval $T_{\mathrm{coh}}$, the normalized optimization latency is further defined as
\begin{equation}
\rho_{\mathrm{P2GD}}=\frac{\tau_{\mathrm{P2GD}}}{T_{\mathrm{coh}}}.
\end{equation}

The proposed method is particularly attractive when a satisfactory beamforming solution can be obtained within a small number of iterations, such that the accumulated iterative signaling overhead remains below or comparable to the pilot cost required for explicit CSI acquisition. This regime becomes increasingly relevant for large SIM dimensions $M$, since the explicit channel-estimation cost grows with the number of unknown spatial dimensions, whereas the per-iteration signaling cost of P2GD is determined by $N$.}

As for the computational complexity, we briefly summarize the dominant costs of the proposed UE-P2GD and BS-P2GD algorithms. In each iteration, the BS first generates the pilot matrix $\mathbf{X}_0 = \mathbf{F}\mathbf{P}^{1/2}$ and updates the effective SIM response $\mathbf{B}_\ell = \mathbf{B}_{\ell-1}\mathbf{H}_{\ell-1}\boldsymbol{\Phi}_\ell$, which requires $\mathcal{O}(L N M^2)$ complex multiplications. The projected-gradient step further involves computing the pseudo-inverse $(\mathbf{B}_\ell^\top)^\dagger$ and multiplying it with the $N\times N$ uplink matrix $\mathbf{U}_0$, which scales as $\mathcal{O}(L N^2 M)$ when $M \geq N$. Since the SIM phase matrices are diagonal, only $M$ entries per layer need to be updated, so the cost of forming and projecting the diagonal gradients from \eqref{BP4} and \eqref{final3}/\eqref{final3_3} is of order $\mathcal{O}(L N M)$ and is negligible compared with the matrix–matrix multiplications. On the UE side, computing the local gradients {\color{blue}$\{\bm{d}_i\}_{i=1}^N$} from \eqref{bp:y} or \eqref{bp_y2} involves only $\mathcal{O}(N^2)$ operations per iteration. Therefore, the overall per-iteration computational complexity of both proposed schemes is
\begin{equation}
\mathcal{O}\big(LN M^2 + LN^2 M\big) \approx \mathcal{O}\big(LNM^2\big).
\end{equation}

Thus, the total complexity over $T$ iterations scales as $\mathcal{O}(T L N M^2)$. Compared with conventional SIM beamforming approaches that require explicit estimation of the full $M\times N$ channel matrix followed by alternating optimization over dense precoding variables, the proposed over-the-air gradient framework avoids cubic dependence on the SIM size and remains computationally tractable even for large-aperture metasurfaces.

\subsection{Conclusion Remark}
{\color{blue}The above analysis characterizes the projected-gradient mechanism under an idealized asymptotic regime with sufficiently small step sizes. The practical objective of P2GD, however, is not to pursue asymptotic convergence through a large number of iterations. Since each iteration incurs additional downlink and uplink signaling overhead, excessive iterations would eventually diminish the advantage of avoiding explicit channel estimation. Therefore, in practical operation, P2GD is executed with a limited iteration budget $T_{\max}$ and may employ a more aggressive step size to obtain a satisfactory beamforming solution within the available coherence and signaling budget.}
\section{Simulation Results}
\subsection{Simulation Scenario}
In this paper, we consider the scenario introduced in Section II, where the simulation parameters are listed in Table \ref{tab:exp}. {\color{blue}All reported results are averaged over 10,000 realizations under each parameter setting. In Fig.~\ref{fig:method}, the error bars extend by 0.1 times the empirical standard deviation above and below the mean. They show scaled realization-to-realization variability, rather than standard errors or confidence intervals. The larger sample size reduces sampling fluctuations in the average curves; it does not imply that every individual realization follows a smooth trajectory.}

{\color{blue}The feedback scaling parameter is initialized as $\alpha_0=0.3$ and updated according to
\begin{equation}
\alpha_{t+1}=\mathrm{clip}\!\left[
0.3\left(\frac{0.01}{\max(E_t^{\mathrm{back}},10^{-12})}\right)^{1/2},
10^{-4},10\right].
\end{equation}
Here, $E_t^{\mathrm{back}}=\|\mathbf U_0^{(t)}\|_F^2/N^2$ is the mean squared magnitude of the BS-received backward signal over the $N$ antennas and $N$ feedback symbols, after feedback limiting, reciprocal propagation, and uplink noise addition, but before division by $\alpha_t$. The operator $\mathrm{clip}$ restricts its argument to $[10^{-4},10]$. No additional temporal smoothing is applied to this update.}

{\color{blue}The default noise power is $q=-150$ dBm per received complex sample in each direction, with independent uplink and downlink noise. In physical units, $\sigma_{\mathrm{phys}}^2=10^{(q-30)/10}$ W. For numerical conditioning, the channel amplitudes are multiplied by $S=10^8$, and the noise variance used in optimization and rate evaluation is scaled consistently as $\sigma_{\mathrm{code}}^2=S^2\sigma_{\mathrm{phys}}^2$. Thus, the default code-domain noise variance is $10^{-2}$. The numerical implementation adds $\epsilon_{\mathrm{ZF}}=10^{-6}$ to the powers entering the logarithmic objectives and their corresponding gradient denominators, including the main signal-power term. This constant is expressed in the scaled power units and is not a pseudoinverse threshold.

The learning-rate parameter controls two fixed step sizes, $\eta_P=\eta$ and $\eta_\Phi=\sqrt{\eta}$, with default $\eta=500$. No line search or gradient-norm normalization is used. After a power update, entries are clipped to at least $10^{-3}$ before rescaling to the total power budget; this is not a strict lower bound after rescaling. The SIM coefficients are projected onto the complex unit disk.

The penalty coefficient is adaptive, with $\lambda_0=10$ and default energy-ratio target $r=1$. Let $E_{\mathrm{main},t}$ and $E_{\mathrm{pen},t}$ denote the mean squared magnitudes of the main and unweighted penalty gradients used by the controller. We update
\begin{align}
\lambda_t^\star&=\sqrt{r\frac{\max(E_{\mathrm{main},t},10^{-12})}
{\max(E_{\mathrm{pen},t},10^{-12})}},\\
\lambda_{t+1}&=\mathrm{clip}\!\left(0.5\lambda_t+0.5\lambda_t^\star,10^{-4},10^4\right).
\end{align}
This targets $\lambda_t^2E_{\mathrm{pen},t}/E_{\mathrm{main},t}\approx r$; smoothing and clipping mean that the target need not hold exactly at every iteration. UE-P2GD measures these energies on the main and penalty matrices before feedback construction. BS-P2GD averages the SIM-coefficient gradient energies over elements and then layers, obtaining the unweighted penalty energy from the weighted energy divided by $\lambda_t^2+10^{-12}$. Thus, the two methods use the same ratio target at different gradient locations.}
\begin{table}[t]
\centering
\begin{tabular}{c|c|c}
\hline
Parameter Name / Function & Notation & Value  \\ \hline
Users / BS antennas  & $N$ & 3 \\
SIM elements per layer& $M$ & $8\times8=64$ \\
$\#$ of SIM layers & $L$ & 5 \\
Carrier frequency & $f_c$ & 28 GHz \\
Wavelength & {\color{blue}$\lambda_c$} & 1.07 cm \\
{\color{blue}Element spacing} & {\color{blue}$d_e$} & {\color{blue}$\lambda_c/2$} \\
{\color{blue}Element aperture} & {\color{blue}$A_e$} & {\color{blue}$\lambda_c^2/4$} \\
{\color{blue}Paths per UE} & {\color{blue}$P_{\mathrm{path}}$} & {\color{blue}3} \\
SIM overall thickness& $D$ & {\color{blue}$4.286$ cm ($4\lambda_c$)}\\
BS \& SIM deployment height & $h$ & 10 m \\
UE distribution range & $(d_{r,\min}, d_{r,\max})$& (100, 250) m \\
Path loss reference & $C_0$ & $-35$ dB \\
Path loss exponent & $\mu$ & 3.5 \\
{\color{blue}Noise per sample/direction} & {\color{blue}$q$} & {\color{blue}$-150$ dBm} \\
{\color{blue}Numerical amplitude scale} & {\color{blue}$S$} & {\color{blue}$10^8$} \\
{\color{blue}Log-power regularization} & {\color{blue}$\epsilon_{\mathrm{ZF}}$} & {\color{blue}$10^{-6}$} \\
{\color{blue}Initial penalty coefficient} & {\color{blue}$\lambda_0$} & {\color{blue}10} \\
{\color{blue}Default penalty ratio target} & {\color{blue}$r$} & {\color{blue}1} \\
{\color{blue}Noise-power sweep} & {\color{blue}---} & {\color{blue}$-175{:}5{:}-135$ dBm} \\
BS maximum power & $P_\mr{max}$ & 100 mW \\
UE maximum power & $P_\mr{UE}$ & 50 mW \\
{\color{blue}Power step size} & {\color{blue}$\eta_P$} & {\color{blue}500} \\
{\color{blue}SIM step size} & {\color{blue}$\eta_\Phi$} & {\color{blue}$\sqrt{500}$} \\
{\color{blue}Learning-rate sweep} & {\color{blue}$\eta$} & {\color{blue}$5,50,500,5000$} \\
{\color{blue}Adaptive $\lambda$ target sweep} & {\color{blue}---} & {\color{blue}\tabincell{c}{$0.1,0.3,1,3,$\\$10,30,100$}} \\
{\color{blue}Maximum iterations} & {\color{blue}$T_{\max}$} & {\color{blue}10} \\
{\color{blue}Number of realizations} & {\color{blue}---} & {\color{blue}10,000} \\
{\color{blue}UE-number sweep} & {\color{blue}$N$} & {\color{blue}$2,3,4,5$} \\
{\color{blue}SIM-size sweep} & {\color{blue}$M$} & {\color{blue}$49,64,81$} \\
{\color{blue}Yao baseline training blocks} & {\color{blue}---} & {\color{blue}22} \\
\hline
\end{tabular}
\caption{Basic simulation parameters and parameter sweeps.}\label{tab:exp}
\end{table}
\subsection{Comparison Algorithms}
{\color{blue}We compare six methods: the proposed UE-P2GD and BS-P2GD, their full-CSI gradient references (UE Full-CSI and BS Full-CSI), and two channel-estimation-based baselines (Yao-LS+GD and Yao-RSLS-iso+GD). The full-CSI references use the corresponding UE-side or BS-side loss with access to the full channel gradient. They serve as references for assessing the effect of gradient projection, rather than guaranteed performance upper bounds for finite-iteration non-convex optimization.

The two Yao baselines use least-squares (LS) and isotropic reduced-subspace least-squares (RSLS-iso) channel estimation based on \cite{yao2024channel}, respectively, followed by gradient-based beamforming optimization. Both use 22 training blocks in the reported comparisons. The horizontal axis in Fig.~\ref{fig:method} counts optimization iterations, each of which includes downlink pilot transmission and uplink feedback for P2GD; it does not include the preceding channel-estimation cost of these baselines. Thus, this figure compares optimization trajectories rather than performance under an equal total signaling or computational budget.}

{\color{blue}For the Yao baselines, each training block contains $N=3$ simultaneous-user uplink symbols with orthogonal DFT pilots satisfying $\mathbf Q\mathbf Q^H=N\mathbf I$. Each UE transmits at 50 mW per symbol. Independent random unit-modulus SIM coefficients are used in each block, without training-codebook MSE optimization. Consequently, 22 blocks require 66 uplink MIMO symbols. The LS estimate uses the stacked sensing matrix pseudoinverse. The RSLS-iso estimate additionally projects the observations onto the eigenspace of $\mathbf A\mathbf R_{\mathrm{iso}}\mathbf A^H$, where $\mathbf A\in\mathbb C^{66\times64}$ is the sensing matrix and $[\mathbf R_{\mathrm{iso}}]_{m,\tilde m}=\operatorname{sinc}(2d_{m,\tilde m}/\lambda_c)$, with $\operatorname{sinc}(x)=\sin(\pi x)/(\pi x)$. This isotropic estimator covariance is distinct from the finite-path channel-generation model. Relative thresholds of $10^{-10}$ are used for eigenvalue retention and pseudoinversion. In all 10,000 fixed-noise realizations at this training budget, both the sensing rank and the retained subspace rank are 64.

After estimation, each baseline performs ten full-gradient UE-objective updates on the fixed estimated channel, using the step sizes, adaptive controls, and feasibility operations described above. The local gradient computation does not add over-the-air observation noise. The achieved rates are evaluated on the true channel with the corresponding noise variance. These are estimator-based adaptations of the cited work to our beamforming setting.}
\begin{table}[t]
\centering\color{blue}
\caption{Signaling overhead under the default setting
($N=3$, $M=64$, and $T=10$).}
\label{tab:overhead}
\begin{tabular}{lccc}
\hline
Method
& Downlink
& Uplink
& Explicit CSI
\\
\hline
UE-P2GD
& $NT=30$
& $NT=30$
& No
\\
BS-P2GD
& $NT=30$
& $NT=30$
& No
\\
CSI-based beamforming
& \multicolumn{2}{c}{$T_{\mathrm{CE}}\ge M=64$ pilot symbols}
& Yes
\\
\hline
\end{tabular}
\end{table}
\subsection{\color{blue} Signaling Overhead Comparison}
{\color{blue}
Using the overhead formulation in Section~V-D, we further provide a quantitative comparison under the default system configuration. Specifically, with $N=3$, $M=64$, and $T=10$, both UE-P2GD and BS-P2GD require
\begin{equation}
\mathcal{O}_{\mathrm{DL}}^{\mathrm{P2GD}}
=
\mathcal{O}_{\mathrm{UL}}^{\mathrm{P2GD}}
=
NT
=
30
\end{equation}
MIMO symbols in the downlink and uplink, respectively.

For explicit CSI acquisition, the ideal identifiability lower bound is $T_{\mathrm{CE}}\ge M=64$ MIMO pilot symbols. This lower bound does not correspond to a specific channel estimator and represents the minimum pilot dimension required to identify the $MN$ unknown complex coefficients of $\mathbf G\in\mathbb C^{M\times N}$.

The resulting signaling requirements are summarized in Table~\ref{tab:overhead}. The implemented 22-block Yao baselines use 66 uplink training symbols, compared with 30 uplink and 30 downlink symbols for ten P2GD iterations. Their local optimization costs are different; equal iteration counts do not imply equal computational time. The comparison shows that, under the adopted finite-iteration setting, the proposed methods operate with a small per-direction signaling budget while avoiding explicit CSI acquisition.
}
\subsection{Comparison Among Different Methods}
\begin{figure}[t]
  \centering
  \includegraphics[width=0.95\linewidth]{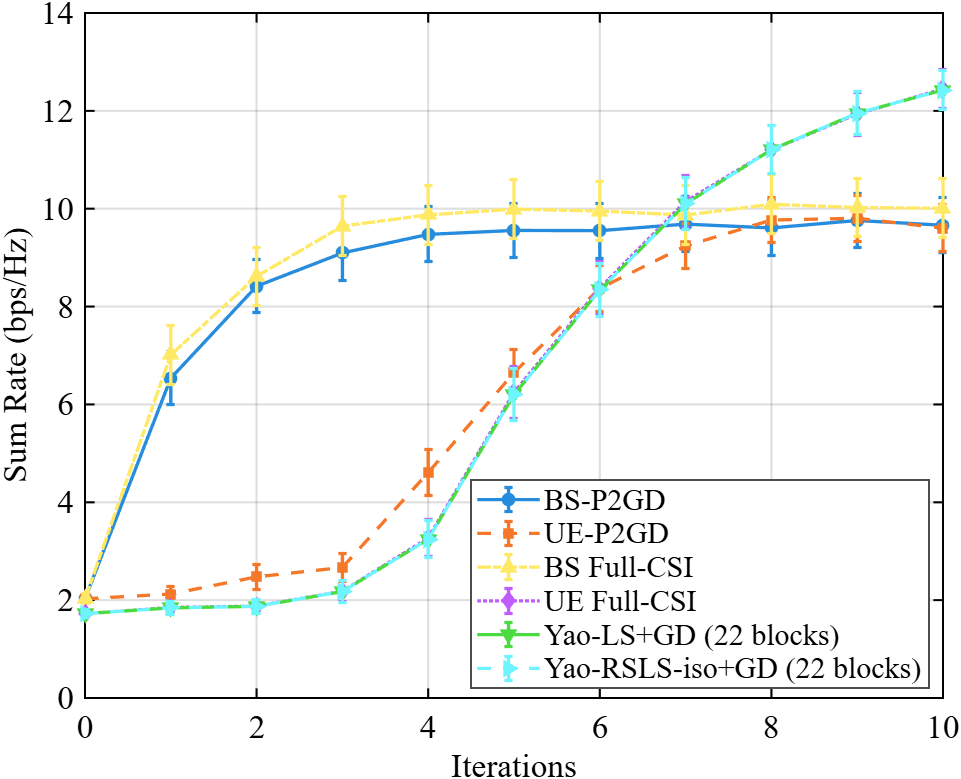}
  \caption{\textcolor{blue}{Performance comparison among different methods, averaged over 10,000 realizations. Error bars show $\pm0.1$ times the empirical standard deviation.}}
    \label{fig:method}
\end{figure}
{\color{blue}Fig.~\ref{fig:method} compares the average sum-rate trajectories of the six methods. BS-P2GD improves rapidly during the first few iterations and then remains close to the BS Full-CSI reference, with a modest projection-related performance gap. UE-P2GD improves more slowly initially and approaches BS-P2GD near the end of the ten-iteration budget. In contrast, UE Full-CSI and the two Yao baselines continue improving and achieve higher average sum rates in the later iterations. The Yao-LS+GD and Yao-RSLS-iso+GD curves nearly overlap with UE Full-CSI under the reported 22-block training setting.

The averages over 10,000 realizations exhibit clear trends with limited late-iteration fluctuations. These results support the usefulness of BS-P2GD for obtaining a satisfactory solution within a small number of over-the-air iterations, while also showing the performance tradeoff relative to explicit-CSI-based optimization. The scaled error bars describe variability across realizations and should not be interpreted as guarantees of stability for individual runs.}

\subsection{Effect of Learning Rate and Iteration Steps}
\begin{figure}[t]
  \centering
  \includegraphics[width=0.8\linewidth]{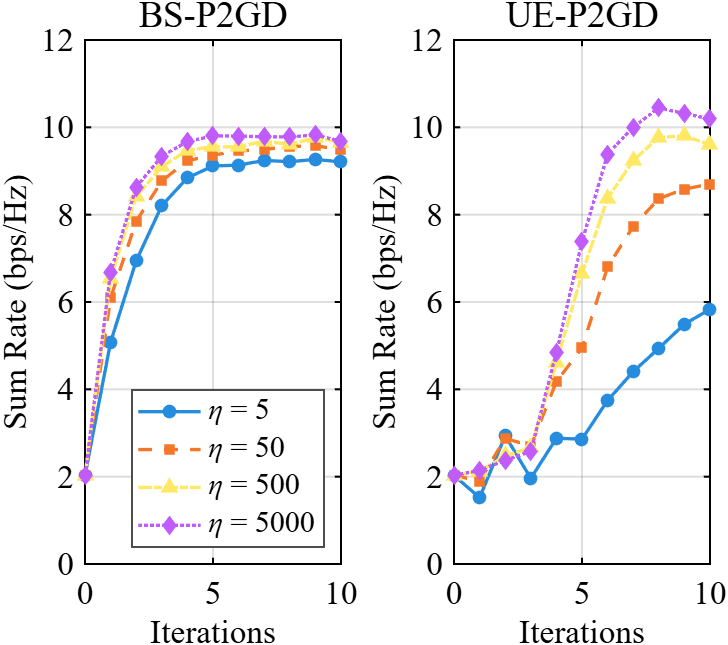}
  \caption{Effect of learning rate in P2GD algorithms.}
    \label{fig:lr}
\end{figure}
Fig.~\ref{fig:lr} compares learning rates $\eta\in\{5,50,500,5000\}$ within a ten-iteration budget. For BS-P2GD, all tested settings improve rapidly, while larger learning rates generally yield higher average rates within this budget. UE-P2GD is more sensitive to the step size: a small learning rate leads to slower improvement, whereas larger values yield substantially higher rates by the final iteration. The largest tested value also produces a small late-iteration decline for UE-P2GD. These observations concern finite-iteration performance rather than asymptotic convergence, and illustrate the importance of choosing a step size appropriate to the available signaling budget.
\subsection{Effect of Background Noise Power}
\begin{figure}[t]
  \centering
  \includegraphics[width=0.8\linewidth]{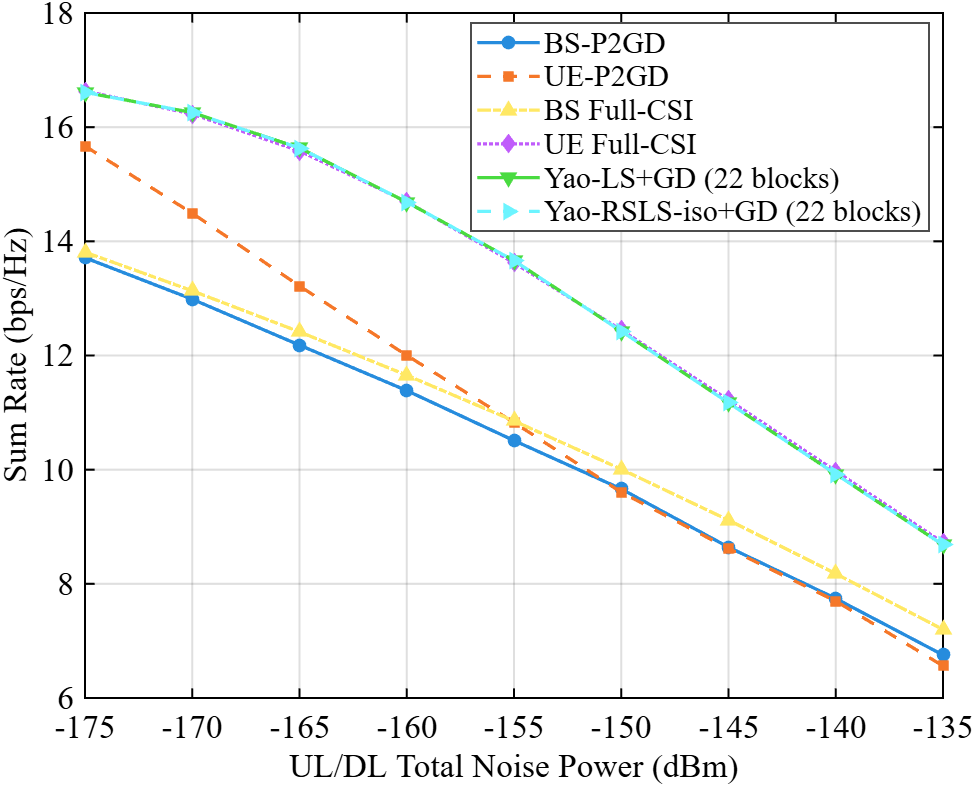}
  \caption{Performance over different background noise levels.}
    \label{fig:noise}
\end{figure}
Fig.~\ref{fig:noise} examines noise powers per received complex sample in each direction from $-175$ to $-135$ dBm in steps of 5 dB. The average sum rate decreases with increasing noise power for all six methods. BS-P2GD remains close to BS Full-CSI across the sweep. UE-P2GD performs better than BS-P2GD at the lower noise levels, while their rates become similar as the noise increases. UE Full-CSI and the two Yao baselines exhibit nearly overlapping curves and achieve higher rates throughout the tested range. Noise affects both the achievable communication rate and the gradient information used by P2GD; this comparison alone does not isolate the contributions of these two effects.

\subsection{Effect of SIM Elements per Layer and UE Numbers}
\begin{figure}[t]
  \centering
  \includegraphics[width=0.8\linewidth]{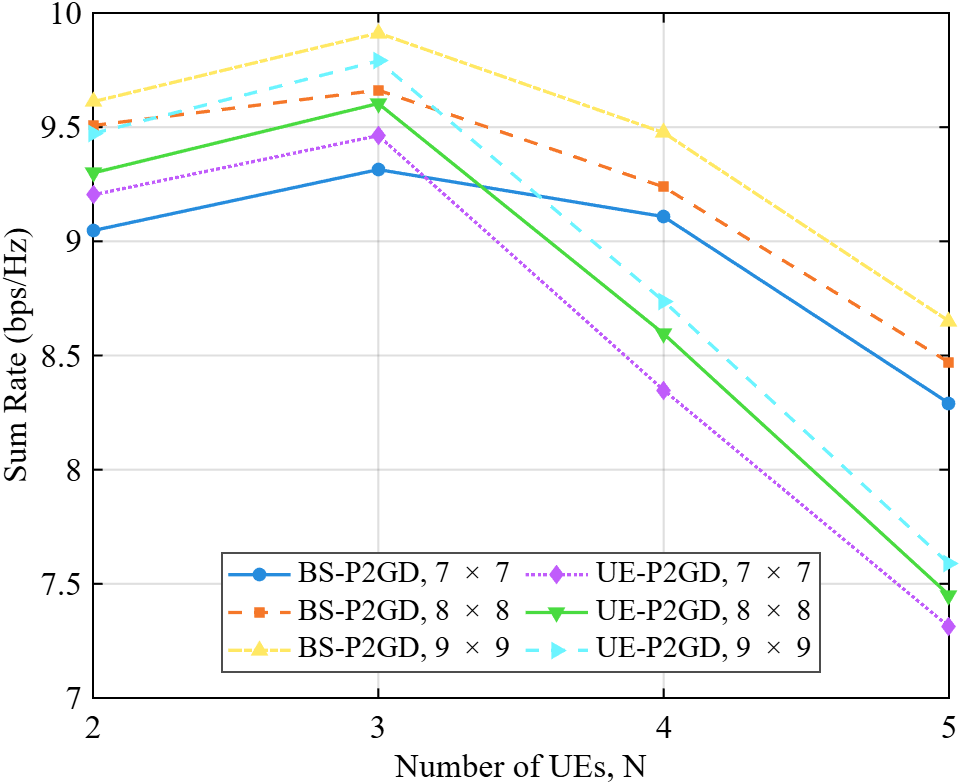}
  \caption{Effect of problem size to the proposed P2GD algorithms.}
    \label{fig:size}
\end{figure}
Fig.~\ref{fig:size} compares $7\times7$, $8\times8$, and $9\times9$ SIM arrays, corresponding to $M\in\{49,64,81\}$, as the number of UEs $N$ varies from 2 to 5. At a fixed UE count, increasing the SIM size generally improves the average sum rate, with only a small difference between the $8\times8$ and $9\times9$ UE-P2GD results at $N=2$. For each array size, the average rate peaks at $N=3$ and decreases as more users are served. This decrease is more pronounced for UE-P2GD, whereas BS-P2GD performs better at $N=4$ and $N=5$. These trends describe the achieved rates of the proposed algorithms under the tested settings, rather than the optimal system sum rate.

\subsection{\color{blue}Sensitivity to the Adaptive $\lambda$ Ratio Target}
\begin{figure}[t]
  \centering
  \includegraphics[width=0.8\linewidth]{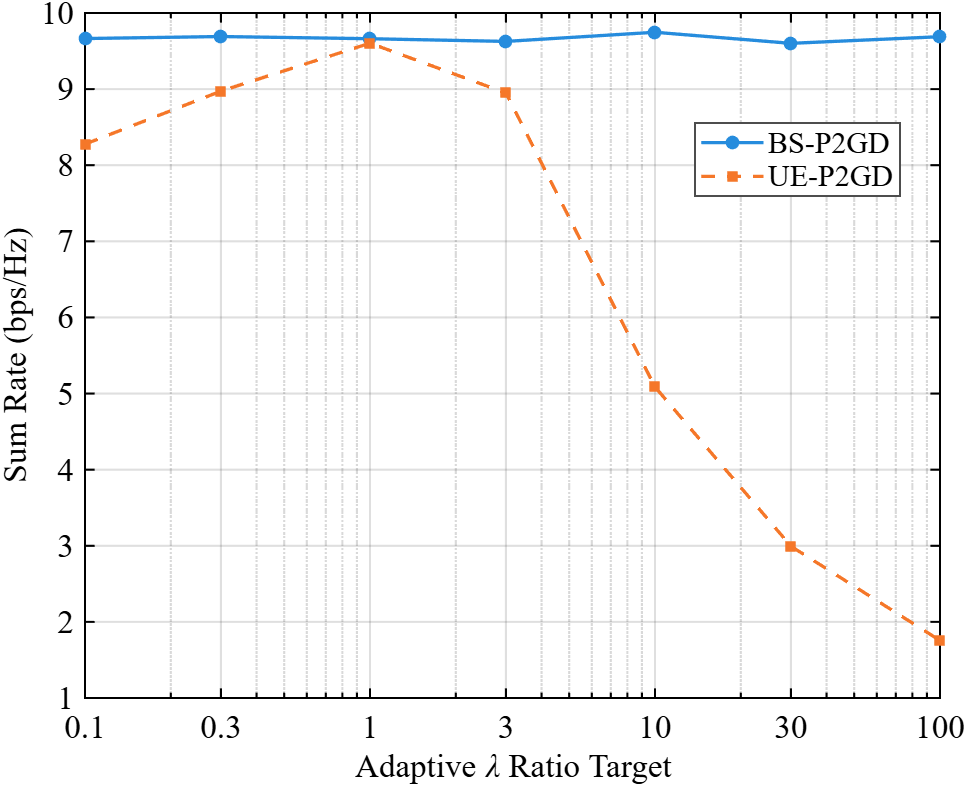}
  \caption{\textcolor{blue}{Average sum rate versus the adaptive $\lambda$ ratio target.}}
  \label{fig:lambda}
\end{figure}
{\color{blue}Fig.~\ref{fig:lambda} examines the adaptive $\lambda$ ratio target over $\{0.1,0.3,1,3,10,30,100\}$. The horizontal axis denotes the target used by the adaptive penalty setting, rather than a fixed value of the penalty coefficient throughout optimization. BS-P2GD maintains a similar average sum rate across the tested range. UE-P2GD is more sensitive: its rate increases up to a target of approximately 1 and then decreases, with a pronounced loss for targets of 10 and above. Thus, BS-P2GD is less sensitive to this tuning parameter in the tested setting, whereas UE-P2GD requires a more careful choice. Table~\ref{tab:lambda_metrics} complements the figure with residual interference and minimum-user-rate results after ten updates.}

{\color{blue}For evaluation, define the physical effective channel including power allocation as $H_{ij}^{\mathrm{eff}}=\sqrt{p_j}\bm g_i^\top\mathbf B^\top\bm e_j$. The residual interference is $I_{\mathrm{res}}=\sum_i\sum_{j\ne i}|H_{ij}^{\mathrm{eff}}|^2$ in watts, and the minimum user rate is
\[
R_{\min}=\min_i\log_2\!\left(1+\frac{|H_{ii}^{\mathrm{eff}}|^2}
{\sum_{j\ne i}|H_{ij}^{\mathrm{eff}}|^2+\sigma_{\mathrm{phys}}^2}\right).
\]
For UE-P2GD, increasing $r$ from 0.1 to 1 improves sum rate and reduces residual interference, while further increases eventually sacrifice sum rate and do not monotonically reduce interference. BS-P2GD has a comparatively insensitive average sum rate across the tested targets, but its residual interference remains nonzero. The small minimum-user rates also show that maximizing this surrogate does not guarantee user fairness. BS-P2GD is attractive when fast initial improvement and reduced tuning sensitivity are priorities; UE-P2GD can be preferable for achieved sum rate in the low-noise settings of Fig.~\ref{fig:noise}, provided its penalty target is chosen appropriately.}

\begin{table}[t]
\centering\color{blue}\footnotesize
\caption{Metrics after ten updates, averaged over 10,000 realizations. Rates are in bps/Hz and interference is in units of $10^{-13}$ W.}
\label{tab:lambda_metrics}
\begin{tabular}{ccrrr}
\hline
Method & $r$ & Sum rate & $I_{\mathrm{res}}$ & $R_{\min}$\\
\hline
UE-P2GD & 0.1 & 8.276 & 0.817 & 0.032 \\
BS-P2GD & 0.1 & 9.664 & 1.062 & 0.037 \\
UE-P2GD & 0.3 & 8.974 & 0.787 & 0.024 \\
BS-P2GD & 0.3 & 9.691 & 1.081 & 0.036 \\
UE-P2GD & 1 & 9.604 & 0.672 & 0.017 \\
BS-P2GD & 1 & 9.662 & 1.061 & 0.036 \\
UE-P2GD & 3 & 8.949 & 0.474 & 0.035 \\
BS-P2GD & 3 & 9.627 & 1.082 & 0.035 \\
UE-P2GD & 10 & 5.087 & 0.461 & 0.150 \\
BS-P2GD & 10 & 9.747 & 1.094 & 0.036 \\
UE-P2GD & 30 & 2.994 & 0.924 & 0.163 \\
BS-P2GD & 30 & 9.601 & 1.130 & 0.036 \\
UE-P2GD & 100 & 1.755 & 1.747 & 0.129 \\
BS-P2GD & 100 & 9.689 & 1.144 & 0.037 \\
\hline
\end{tabular}
\end{table}

\section{Conclusion and Future Directions}

This work introduces a gradient-driven optimization framework for SIM-assisted TDD MIMO systems, aiming to overcome the inherent limitations of conventional channel estimation and precoding strategies. By exploiting channel reciprocity and incorporating over-the-air gradient feedback, the proposed approach enables the joint refinement of SIM configuration and transmit power allocation under realistic system constraints. {\color{blue}The theoretical analysis characterizes the projected-gradient mechanism under idealized conditions and provides insights into its behavior, while the performance of the practical non-convex algorithms is evaluated numerically.} 
{\color{blue}Simulations show that BS-P2GD remains close to its full-CSI reference within ten over-the-air iterations, whereas explicit-CSI-based optimization attains higher sum rates at a larger pilot cost. The proposed approach is therefore most attractive when frequent channel variations make repeated explicit channel estimation costly.}

{\color{blue}The surrogate loss functions and finite-iteration parameter choices still influence the achieved sum rate, especially for UE-P2GD. Improving their alignment with the sum-rate objective remains a direction for further investigation.}
Meanwhile, the modular nature of the proposed architecture suggests strong extensibility, allowing integration with advanced optimization techniques and adaptation to emerging paradigms such as multi-base station coordination, UAV-assisted networks, and semantic communication. These attributes position the framework as a versatile foundation for future research on intelligent and scalable wireless systems.
\printbibliography
\end{document}